\documentclass[aps,pra,twocolumn,showpacs,floatfix]{revtex4-2}

\usepackage[utf8]{inputenc} %useful to type directly diacritic characters
\usepackage{amssymb,amsmath}
\usepackage{graphicx}
\usepackage{braket}
\usepackage{siunitx}
\usepackage{xcolor}
\usepackage[normalem]{ulem}
\usepackage{bm}

\def\Rb87{^{87}\mathrm{Rb}}                     % Rb 87
\def\ex{\mathbf{e}_x}

\def\Isat{I_\mathrm{sat}}
\def\Ibar{I/\Isat} % \def\Ibar{{\bar I}}
\def\dbar{\delta / \Gamma}
\def\tbar{{\Gamma t_{\rm m}}}
\def\tm{t_{\rm m}}

\begin{document}

%TC:ignore
\title{Repeated weak measurements: watching quantum correlations evolve}

\author{Emine~Altunta\c{s}}
\email{altuntas@ou.edu}
\affiliation{Joint Quantum Institute, National Institute of Standards and Technology, and University of Maryland, Gaithersburg, Maryland, 20899, USA}
\author{I.~B.~Spielman}
\email{ian.spielman@nist.gov}
\affiliation{Joint Quantum Institute, National Institute of Standards and Technology, and University of Maryland, Gaithersburg, Maryland, 20899, USA}
\homepage{http://ultracold.jqi.umd.edu}
\date{\today}

\begin{abstract} 
Experimental access to many-body quantum systems is often limited by measurement backaction, and key dynamical properties are typically obtained by perturbing a system and measuring its response.
Here we replace this active paradigm with a minimally invasive protocol based on a pair of weak quantum measurements that leverages measurement backaction as a strength.
By correlating time-separated measurements with the first detecting fluctuations---of any sort---and the second tracking their time evolution, our method directly measures dynamical correlation functions without external perturbation. 
We demonstrate this technique in an atomic Bose–Einstein condensate using phase-contrast imaging to obtain the two-time density-density correlation function known as the Van Hove function and, through its Fourier transform, the dynamical structure factor.
Due to the role of spatial correlations in scattering, these quantities underpin neutron and X-ray scattering and atomic Bragg spectroscopy.
This approach is broadly applicable, providing access to correlation functions between any pair of observables amenable to weak measurement, thereby going beyond the capabilities of conventional strong measurements. 
We further isolate the role of quantum backaction through Aharonov’s post-selection-based quantum weak values.
\end{abstract}

\maketitle

%TC:endignore

%----------------------------------------------------------------------------------------
%%%%%%%%%%%%%%%%%%%%%%%%%%    INTRODUCTION    %%%%%%%%%%%%%%%%%%%%%%%%%%
%----------------------------------------------------------------------------------------

\noindent Randomness and uncertainty---hallmarks of quantum mechanics---emerge primarily at the interface between the quantum and classical worlds, where measurement plays a pivotal role.
Quantum projection noise, accompanied by backaction, arises when a classical observer measures a quantum system.  
In the case of a standard strong measurement, this fully collapses the wavefunction into a state consistent with the observation, erasing all memory of the initial state and preventing subsequent measurements from obtaining new information.
Von Neumann's generalized theory of quantum measurement\cite{Neumann2018} extends this description to include weak measurements that provide reduced information and consequently only partly collapse the wavefunction. 
Therefore, a single system undergoing time evolution can be sequentially interrogated, with each weak observation providing new information including the impact of prior measurements' backaction.
Weak and partial measurements underpin feedback-quantum-control\cite{Rossi2018}, inform approaches to quantum error correction\cite{Terhal2015}, and are increasingly used in quantum sensing and precision measurement.\cite{Hosten2008,Zhang2015,Greve2022}
In ultracold atomic gases, they enable real-time measurements of closed-system dynamics including evaporative cooling\cite{Freilich2010,Zeiher2021}, and---via measurement backaction and feedback---the creation of number- and spin-squeezed states in large ensembles\cite{Kristensen2017, Christensen2021, Bohnet2014}.

Here we focus on pairs of time-separated weak measurements of density in atomic Bose--Einstein condensates (BECs) and show that excitations both observed and created by the first measurement leave tell-tale signs that can be detected in a second measurement after a time delay $\delta t$, possibly displaced a distance $\delta x$. 
Cross-correlations between these measurements directly access these signs in terms of the Van Hove function $G(\delta x, \delta t)$, a two-time correlation function.
In condensed matter systems its space--time Fourier transform, the dynamical structure factor (DSF) $S(k,\omega)$ quantifying the many-body excitation spectrum, is often measured by neutron scattering.\cite{Krivoglaz1969,Lovesey_1984}
In quantum gases, the DSF has been measured using Bragg scattering of atoms from a traveling lattice potential\cite{Ozeri2005} that imparts a well-defined momentum $\hbar k$ and energy $\hbar \omega$.
These well-established techniques require a calibrated perturbation applied to an equilibrium system, followed by a suitable measurement.
Our experiment shows that this complexity can be avoided simply by observing the system with a sequence of weak quantum measurements: a many-body analogue of quantum nonlinear spectroscopy\cite{Meinel2022}.

\begin{figure}[tb!]
\includegraphics{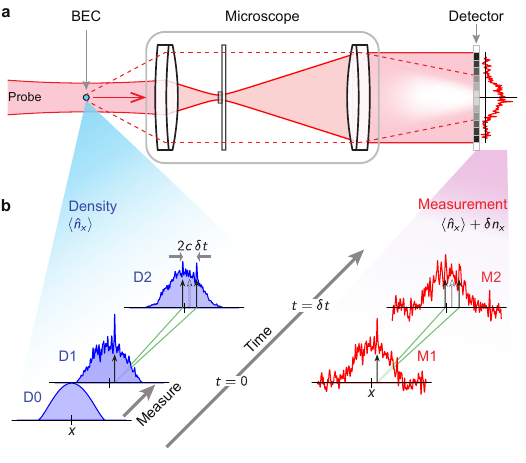}
\caption{{\bfseries Principle of method.} 
{\bfseries a} Schematic of the experimental apparatus.
A far-detuned probe laser (solid red) is phase-shifted as it traverses a BEC (blue) and is partly scattered (dashed red). 
A phase dot in the Fourier plane shifts the unscattered light's optical phase by $\pi/2$ before the resulting intensity is imaged.
{\bfseries b} Conceptual measurement principle.
Two measurement outcomes are shown in red: M1, a measurement of the initial state at $t=0$; and M2, after an evolution time $\delta t$.
Three density profiles are shown in blue: D0, that of the initial state; D1 just following M1, with measurement-induced excitations correlated to the noise in M1; and D2 just prior to M2.
A representative initial fluctuation at $t=0$ (vertical black arrow) evolves into a pair of correlated features spaced by $2 c \delta t$. 
Solid green lines schematically indicate the trajectory of correlations spreading at the speed of sound. 
}
\label{Fig1:Method}
\end{figure}

Phase-contrast imaging (PCI), dark-field imaging, and Faraday imaging are notable examples of weak-measurement methods for cold-atom systems\cite{Andrews1996,Ketterle1999,Gajdacz2013}.
We employ PCI to weakly measure the per-pixel atom number (illustrated in Fig.~\ref{Fig1:Method}{\bf a}) on a camera at pixel coordinate $x$, and construct an ensemble by repeating each experiment $128$ times.
This process can be described as shown in Fig.~\ref{Fig1:Method}{\bf b}, where the first measurement outcome (M1) at time $t=0$ is written as $n_{x} = \langle \hat n_{x}\rangle + \delta n_{x}$ with contributions from the expectation value of the atom number $\langle \hat n_{x}\rangle$ just prior to the measurement (D0) and its quantum projection noise 
\begin{align}
\delta n_{x} &= \frac{m_{x}}{\varphi}. 
\label{eq:meas1}
\end{align} 
The noise amplitude is governed by a measurement strength $\varphi\ll 1$ and the spatial structure is set by a zero-mean random variable $m_x$ (see Supplementary Information Sec.~\ref{app:theory_concept} and below). 
The associated quantum backaction locally updates the many-body wavefunction $\ket{\Psi}$ by an amount $\varphi \sum_x m_{x} \left(\hat n_x -  \langle \hat n_{x} \rangle \right)\ket{\Psi}$, pulling the initial density profile D0 closer to the measurement outcome M1, and thereby creating additional excitations in the post-measurement density profile (D1).\cite{Plenio1998,Wiseman2011,KrausFootnote}

A second measurement (M2) of the same system after a time delay $\delta t$ has measurement noise
\begin{align}
\delta n'_{x + \delta x} &=  \frac{m'_{x + \delta x}}{\varphi'} + \varphi \sum_{x} m_x \langle \{\delta\hat n_{x}(0), \delta\hat n_{x + \delta x}(\delta t)\}\rangle, \label{eq:meas2}
\end{align}
at pixel $x + \delta x$ (primes identify second-measurement quantities).
In addition to its own projection noise, this expression carries the imprint of the first measurement's backaction through the Van Hove function's real part $G_{\delta x}(\delta t) = N^{-1}\sum_x \langle \left\{ \delta \hat  n_{x}(0), \delta \hat n_{x+\delta x}(\delta t)\right\}\rangle/2$ for $N$ atoms.
% Here is a reference for this including the N^-1: J. P. Hansen and I. R. McDonald, Theory of Simple Liquids (London: Academic Press, 2006).
Intuitively, the first measurement both provides information about existing excitations and creates additional excitations by quantum backaction; the second measurement then reveals how these have evolved in time.
This result does not depend on a-priori knowledge of the wavefunction or even the assumption of a pure state (see Supplementary Information Sec.~\ref{app:theory_concept}).
Our experiments began by obtaining $G_{\delta x}(\delta t)$ as the cross-correlation function (CCF) of the two measurements, independent of their strength.
Then, inspired by Aharonov’s quantum weak values (QWV),\cite{Aharonov1988} we post-select second-measurement outcomes based on the first, to obtain a QWV $\propto \varphi G_{\delta x}(\delta t)$, thereby exposing the specific role of quantum backaction encoded in $\varphi$.

%----------------------------------------------------------------------------------------
%%%%%%%%%%%%%%%%%%%%%%%%%%    EXPERIMENT    %%%%%%%%%%%%%%%%%%%%%%%%%%
%----------------------------------------------------------------------------------------
%\subsubsection*{Experimental system}\label{sec:ExpOverview}

Our experimental sequence, schematically shown in Fig.~\ref{Fig1:Method}{\bf b}, began with $\Rb87$ BECs prepared in the $\ket{f=2, m_F=2}$ hyperfine state with $\approx 2.0\times10^5$ atoms confined in a crossed dipole trap. 
We studied dynamics along the BEC's $\approx 90~\mu{\rm m}$ long-axis by imaging the atomic density using PCI at time $t=0$ and again after a delay $\delta t$, yielding measurements M1 and M2.
During each measurement, the BEC was illuminated for a duration $t_m = 16.4~\mu\rm{s}$~\footnote{To mitigate parasitic excitations that occur while the probe is applied, each measurement consisted of a pair of pulses of duration $t_m/2$ separated by a well chosen small delay.\cite{Altuntas2023a}} and, for each set of experimental parameters, we repeated this process $128$ times.
We then computed ensemble-averaged CCFs or post-selected QWVs from these ensembles (see Supplementary Information Sec.~\ref{app:analysis}).

The off-resonant probe laser was red-detuned by $\dbar = -120$ to $-150$ linewidths and had an intensity up to $\Ibar = 12$, where $\Gamma$ is the natural linewidth and $\Isat = 1.67\ {\rm mW}/{\rm cm}^2$ is the saturation intensity. 
Under these conditions, the atoms interact weakly with the probe via a predominantly dispersive susceptibility, which locally phase-shifts the probe light by an amount proportional to the atomic density.
The PCI microscope (Fig.~\ref{Fig1:Method}{\bf a}) is an on-axis homodyne interferometer that converts this phase shift at the object plane into a measurable change in intensity at the image plane.\cite{Zernike1942} 
This yields the number of atoms per-pixel, integrated along the imaging axis.
The dimensionless measurement strength $g = (\tbar)^{1/2} (\Ibar)^{1/2} / (\dbar)$ governs the signal-to-noise ratio (SNR) scaling of these measurements\cite{Altuntas2023a,Altuntas2023}.
For example, the mean number of photons scattered per atom is $N_{\rm scat} = g^2 / 8$, and its associated photon shot noise scales as $\sqrt{N_{\rm scat}}\propto g$.
For a PCI imaging system (as in Fig.~\ref{Fig1:Method}{\bf a}) with numerical aperture (NA) set by the first lens, the effective measurement strength $\varphi \sim {\rm NA}\times g$ (see Supplementary Information Sec.~\ref{app:strength}) scales inversely with the resolution limited spot-size.\footnote{As discussed in the methods, the largest resolution-limited measurement strength in our experiment is $\varphi\approx0.1$, at $g=1.0$.}

%----------------------------------------------------------------------------------------
%\subsubsection*{Dynamical structure factor from correlations}\label{sec:MainResult}

\begin{figure}[tb!]
\includegraphics{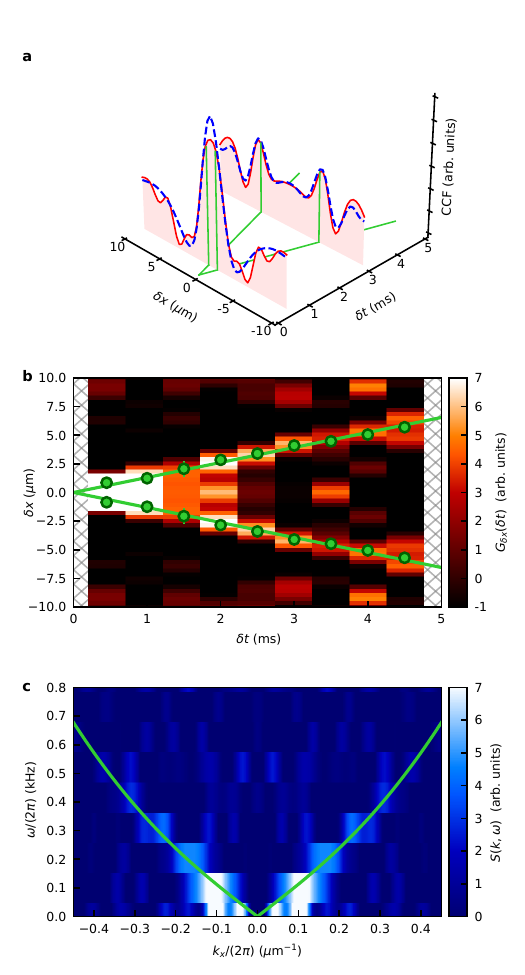}
\caption{{\bfseries Van Hove function and dynamical structure factor.} 
All measurements were taken at $\dbar = 124.3$ and averaged over 128 repetitions of the experiment.
M1 had $g = 0.3$ while M2 used $g' = 1.0$, corresponding to $N_{\rm scat} \approx 0.01$  and $\approx 0.125$, respectively. 
{\bfseries a} Experimental (solid red) and modeled (dashed blue) CCFs at $\delta t = 0.45~\rm{ms}$ and $\delta t = 3.0~\rm{ms}$ are shown along with a solid green line marking the trajectory of excitations traveling at the sound speed $c$.
The extra oscillatory structures in the short-time data reflect typical statistical fluctuations.
{\bfseries b} Symmetrized Van Hove function. 
Individual green circles locate the peak displacement obtained from individual fits to the modeled line-shape as in {\bfseries a} (see Supplementary Information Sect.~\ref{app:lineshape_model}), while the solid green line plots $c t$ obtained from a single global fit to the complete Van Hove dataset.
{\bfseries c} Dynamical structure factor computed from {\bfseries b}.
The solid green curve plots the Bogoliubov dispersion corresponding to the speed of sound obtained in {\bfseries b}. 
The error bars denote the standard error of the mean (s.e.m.).
}
\label{Fig2:2FK_CCF}
\end{figure}

With this experimental framework, we extract the Van Hove function from the second measurement's noise using our first approach: the cross-correlation analysis of density fluctuations. 
Because fluctuations present in M1 are echoed in M2, the ensemble-averaged (denoted by an overline) two-time noise correlation function $\sum_x\overline{\delta n_{x}(0) \delta n_{x+\delta {x}}(\delta t)}$ directly yields $G_{\delta x}(\delta t)$, independent of measurement strength.
Figure~\ref{Fig2:2FK_CCF}{\bf a} shows a pair of CCFs obtained with this procedure.
The first CCF, recorded at $\delta t = 0.45~\rm{ms}$, exhibits a single imaging-resolution-limited peak.
By contrast, the CCF at $\delta t = 3.0~\rm{ms}$ develops a pair of peaks at $\delta x \approx \pm 5\ \mu{\rm m}$, directly revealing the propagation of correlated density fluctuations.

Small amplitude density excitations obey the Bogoliubov dispersion\cite{Bogoliubov1947}, with an independently measured\cite{MuFootnote} long-wavelength speed of sound $c = 1.35(10)\ \rm{mm/s}$.
With this context, the correlation signal for $\delta t = 0.45~\rm{ms}$ remains near $\delta x = 0$ because these excitations propagate only $\approx 0.6\ \mu{\rm m}$, well below our $\approx 2\ \mu{\rm m}$ imaging resolution (${\rm NA} = 0.32$).
By $3.0~\rm{ms}$, Bogoliubov excitations become resolvable, with $c\,\delta t$ consistent with the observed displacement. 
The dashed curves, obtained from a local-density-approximation (LDA) model of Bogoliubov excitations viewed through our imaging system\cite{Altuntas2021}, capture the observed lineshape.

We assemble a family of such experimental curves into the Van Hove function $G_{\delta x}(\delta t)$ in Fig.~\ref{Fig2:2FK_CCF}{\bf b}; this reveals a sharp ridge corresponding to a correlation peak moving linearly in time. 
From a global fit of the LDA imaging model to the full dataset, we determine the maximum longitudinal speed of sound $c = 1.31(2)\ \rm{mm/s}$; this agrees with the independent measurement noted above. 
The green line plots the ballistic motion $c\delta t$ deduced from this fit.

Finally, Fig.~\ref{Fig2:2FK_CCF}{\bf c} plots the DSF obtained from the two-dimensional Fourier transform of the measured Van Hove function.   
This directly maps the system's collective excitation spectrum as a function of wavevector $k$ and angular frequency $\omega$.
The solid green curve shows the Bogoliubov dispersion relation, computed using the speed of sound obtained in {\bf b}, closely aligning with locations of maximum spectral weight.
The maximum observable wavevector is NA-limited to $k_{\rm NA}/(2\pi) = 0.41\ \mu{\rm m}^{-1}$; this verifies that our observations are confined to long-wavelength, linearly dispersing phonons.
Although these data were taken with deeply degenerate BECs with condensate fraction $R_c \approx 0.98$ and temperature $T \approx 20~\mathrm{nK}$, the corresponding thermal energy scale $k_{\rm B}T \approx h\times 400~\mathrm{Hz}$ lies within the DSF's measurement window.
This suggests that the observed phonon modes had significant thermal occupation below this value, contributing enhanced low-$\omega$ weight.

%--------------------------------------------------------------------------------
%\subsubsection*{Weak values}\label{sec:WeakValues}

In 1988, Aharonov and colleagues introduced a protocol in which a weak quantum measurement is post-selected on the outcome of a subsequent projective readout measurement.\cite{Aharonov1988,Dressel2014}
The resulting QWV can lie outside the eigenvalue spectrum of the weakly measured observable, thereby amplifying weak-measurement signals.
Here, we adapt this idea to our setting by replacing the projective readout with a second weak measurement, yielding a minimally destructive protocol that preserves the QWV amplification mechanism.

Our selection protocol retains the second measurement's noise $\delta n'_{x + \delta x}$ at pixel $x + \delta x$ only when the corresponding first measurement at pixel $x$ satisfies $\delta n_x>0$, yielding the conditional noise signal $\delta n^{+}_{x + \delta x}$.
This post-selection discards a fraction $f^+_d\approx0.5$ of the data.
Averaging over first-measurement pixels $x$ and experimental repetitions yields a weak value $\overline{\delta n^{+}_{\delta x}} = \varphi G_{\delta x}(\delta t) / \sqrt{\pi}$, proportional to the Van Hove function; laboratory noise attenuates this expression by a constant factor (see Supplementary Information Sect.~\ref{app:theory_concept}).
Unlike the CCF signal, the QWV amplitude grows linearly with the first-measurement strength $\varphi$,  thereby isolating and highlighting the role of quantum noise.
Applying the same reasoning to the half of the dataset with $\delta n_x < 0$ yields an equal-magnitude signal of opposite sign.
Combining the positive and negative subensembles (i.e., sign-weighting and averaging) allows us to retain the entire dataset (overall discarded fraction $f_d=0$) and thereby reduce statistical uncertainty.

\begin{figure}[tb!]
\includegraphics{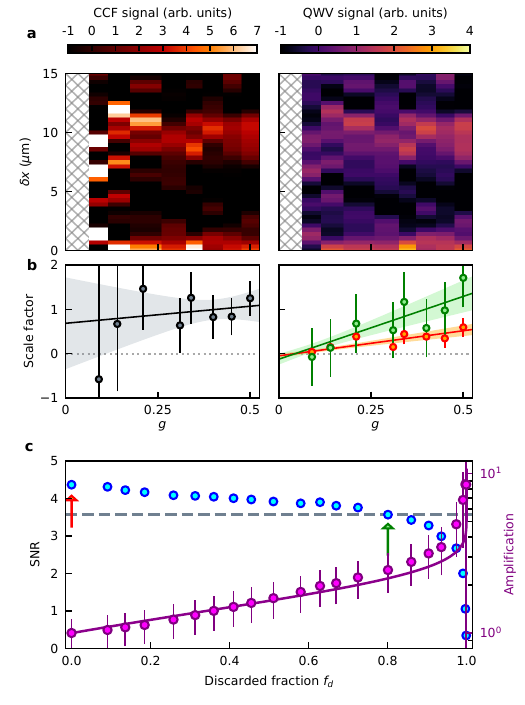}
\caption{
{\bf Quantum weak values.}
{\bf a} Correlation signals from CCF (left) and QWV (right; $f_d=0$) at fixed evolution time $\delta t=8\ {\rm ms}$, plotted versus the first-measurement strength $g$.
{\bf b} Corresponding peak amplitudes: CCF (black) and QWV for $f_d=0$ (red, from {\bf a}) and $f_d=0.8$ (green).
Solid lines are linear fits with single-$\sigma$ confidence bands.  
{\bf c} QWV amplification factor (purple; slope of QWV amplitude versus $g$) and SNR (blue) versus $f_d$.
The dark purple curve is the theoretical amplification (see Supplementary Information Sec.~\ref{app:theory_concept}); the dashed gray curve shows the SNR from the CCF analysis.
The error bars show the s.e.m.
}
\label{Fig:WeakValues}
\end{figure}

Figure~\ref{Fig:WeakValues}{\bf a} compares standard CCFs with QWV measurements across a range of first-measurement strengths $g$, all at a fixed evolution time $\delta t = 8\ {\rm ms}$.
The presence of correlation peaks at $\approx 10 \mu{\rm m}$ in both cases validates our QWV protocol.
In the CCF data, the noise floor falls as $g$ increases, eventually exposing a correlation feature comprising a peak at $\delta x \approx 10\ \mu{\rm m}$ nested between aperture-induced troughs.
By contrast, the QWV noise floor is approximately independent of $g$, and the correlation feature emerges from this background as $g$ increases. 
This impression is confirmed in Fig.~\ref{Fig:WeakValues}{\bf b}, showing the CCF (left) and QWV (right) signal amplitudes as a function of $g$ (the red QWV data corresponds to {\bf a}).

Next, we examine the characteristic QWV amplification by increasing the post-selection threshold from zero, thereby discarding an increasing fraction $f_d$ of the data.
Figure~\ref{Fig:WeakValues}{\bf b} (right) contrasts two cases: in the first (red), all the data is fully retained ($f_d = 0$), whereas in the second (green) a significant fraction of the data is rejected ($f_d=0.8$).
Both are proportional to $g$, and the larger slope for $f_d=0.8$ confirms the expected signal amplification.
We quantify these observations with uncertainty-weighted linear fits of $a(g - b)$, with jointly fit intercept parameter $b$ consistent with zero.
The fits show that the CCF amplitude is independent of $g$ (slope consistent with zero), while the QWV amplitude grows linearly with $g$; moreover, the fitted slope rises with $f_d$, while the intercept remains consistent with zero.

Figure~\ref{Fig:WeakValues}{\bf c} extends this analysis to the full range of discarded fractions.
The purple circles (slopes $a$ obtained from the linear fits with intercept parameter $b$ fixed to zero) confirm that the QWV amplification factor increases monotonically with $f_d$; this trend agrees with the theoretical relation with no free parameters (purple curve, see Supplementary Information Sec.~\ref{app:weak_values}).
The same panel shows that the QWV signal-to-noise ratio (SNR, blue) initially decreases slowly with $f_d$, and then drops rapidly once majority of the data are rejected.
For comparison, the dashed gray line plots the SNR obtained from the standard CCF analysis, i.e., without any weak value processing.

%--------------------------------------------------------------------------------
%\subsubsection*{Temperature dependence}\label{sec:Temperature}

\begin{figure}[tb!]
\includegraphics{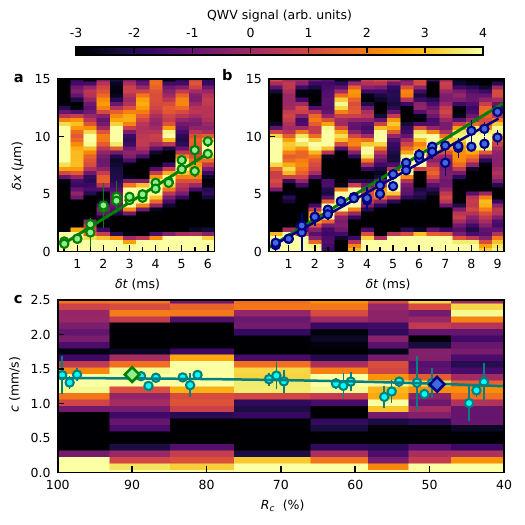}
\caption{
{\bf Temperature dependence at fixed atom number.}
All data used a three measurement series with M1, M2 and M3 each separated by $\delta t$ and were taken at $\dbar = 153.7$, with $g = 0.3$, $g' = 0.5$, and $g'' = 0.5$.
Circles reflect fits to QWVs from individual measurement pairs, while solid curves describe fits to multiple QWVs simultaneously.
Color plots are averages of these QWVs.
QWV signals from {\bfseries a} a deeply-degenerate [$R_c \approx 0.88(2)$] BEC and {\bfseries b} a partly condensed [$R_c = 0.49(2)$] BEC.
Lines show correlations spreading at the speed of sound obtained from the full simultaneous fit. 
The green curve in {\bfseries b} replicates that in {\bfseries a} for reference.
{\bfseries c} QWV signal versus condensate fraction $R_c$.
Solid curve plots the $R_c$-dependent speed of sound described in the text, with the speed of sound at $R_c=1$ obtained from the simultaneous fit.
For each $R_c$, the three individually-determined speeds of sound (circles) are horizontally displaced for clarity.
Diamonds mark the speed of sound determined in {\bf a} and {\bf b}.
Error bars indicate the s.e.m.
}
\label{Fig:SpeedOfSound}
\end{figure}

We conclude by demonstrating the applicability of the QWV protocol to higher-temperature BECs where both the condensate fraction $R_c$ and the sound speed $c$ decrease in tandem.
In a harmonic trap this is weak, consisting of a power-law: $c\propto R_c^{\alpha}$ (with $\alpha = 1/5$, valid until very small $R_c$) with a trap-frequency-dependent proportionality constant (see Supplementary Information Sec.~\ref{app:speed_of_sound}).

Figures~\ref{Fig:SpeedOfSound}{\bf a} and {\bf b} plot the QWV Van Hove function (with $f_d = 0$) for deeply degenerate [$R_c = 0.88(2)$ and $T \lesssim 50\ {\rm nK}$] and partly condensed [$R_c = 0.49(2)$ and $T = 170(10)\ {\rm nK}$] BECs with total atom number held constant at $2.4(3)\times10^5$. 
For improved SNR these data employed a three measurement series with M1, M2 and M3 each separated by $\delta t$.
The color represents an average of both M1-M2 and M2-M3 weak values, while circles are obtained from separate fits (see Supplementary Information Sect.~\ref{app:lineshape_model}) to the individual QWVs.
The solid lines depict the displacement of correlations traveling at a speed of sound determined by a global fit to both weak values---accounting for their $g$-dependent amplification factor---for all $\delta t$.
The data at both temperatures are qualitatively similar, and agree with the expectation that correlations spread slightly more slowly at reduced $R_c$: $c$ falls from $1.42(2)\ {\rm mm/s}$ to $1.28(1)\ {\rm mm/s}$.
For direct comparison, the low-temperature best-fit curve from panel {\bf a} is overlaid as the green curve in panel {\bf b}.

Figure~\ref{Fig:SpeedOfSound}{\bf c} summarizes the sound-speed dependence on condensate fraction (to further enhance SNR, this analysis used all three measurement pairings).
Circles are obtained from fits to individual QWVs, diamonds represent the values determined in {\bf a} and {\bf b}, and the solid curve results from a global fit to the expected form with the $R_c=1$ speed of sound as the only free parameter yielding $1.37(1)\ {\rm mm/s}$.
%2026/3/3 EA confirmation note: this is the global fit result without alpha
This is in agreement with the speed of sound determined in {\bf a}, accounting for a $\approx 30\ \%$ number reduction in {\bf c}. 
Taken together, these results show that our QWV protocol delivers quantitative correlation measurements even in partially condensed, higher-temperature gases.

%-----------------------------------------------------------------------------------
%%%%%%%%%%%%%%%%%%%%%%%%%%    DISCUSSION AND OUTLOOK %%%%%%%%%%%%%%%%%%%%%%%%%%
%-----------------------------------------------------------------------------------
%\subsubsection*{Discussion and outlook} \label{sec:Conc}

Understanding and optimizing quantum measurement is essential for advances in both fundamental physics and quantum technologies.
In this work, we experimentally demonstrate that a pair of time-separated weak measurements in a Bose--Einstein condensate expose dynamical correlations and directly probe the role of projection noise and its backaction. 
By harnessing the interplay between the measurement record and fluctuations, we determined the symmetrized dynamical structure factor without additional drive fields, what could be thought of as a many-body version of quantum nonlinear spectroscopy\cite{Meinel2022}.

Weak, global, spatially resolving measurements of the type employed here stand in contrast to the ``partial measurements'' common in quantum error correction\cite{Terhal2015}, which generally employ strong measurements of a handful of ancilla qubits.
Interleaving partial measurements with unitary evolution can yield measurement-induced phases that exhibit steady-state volume-law entanglement when conditioned on the entire measurement record\cite{Skinner2019,Gullans2020,Noel2022,Koh2023,Hoke2023}.
Our weak-value experiments nevertheless mirror these ingredients: post-selection on the measurement record isolates subensembles with desired properties.
This suggests that global weak measurements can realize analogues of measurement-induced phases in large-scale degenerate gases, without the need for spatial control of the measurements\cite{Szyniszewski2020}.
Furthermore, when these global weak measurements are augmented with real-time feedback, they can stabilize nearly pure, far-from-equilibrium states\cite{Young2021}, making them a promising tool for future quantum technologies.

Our results show that the interplay between measurement, backaction, and unitary evolution can reveal---and potentially engineer---dynamical phenomena in interacting quantum systems.
Because this technique relies solely on repeated weak measurements, it is broadly applicable across quantum platforms, including arrays of superconducting qubits, trapped-ion systems, and photonic lattices.
Any pairwise combination of weakly observable quantities (e.g., magnetization, spin-density, even momentum) can be probed in this way to obtain the corresponding two-time correlation function.
Furthermore, this method applies to all quantum states---low-energy, far-from-equilibrium, pure or mixed---not just those near thermal equilibrium\cite{Schuckert2020}.
Indeed, the unequal time correlation function in Eq.~\eqref{eq:meas2}, appears in the Keldysh formalism of non-equilibrium field theories as a type of four-field Greens function\cite{kamenev2005}.
Looking ahead, harnessing measurement-based control may enable on-the-fly cooling protocols and measurement-driven quantum computing, pointing to a route for exploring out-of-equilibrium dynamics and stabilizing exotic many-body phases.\\

%TC:ignore

\noindent {\bf Acknowledgments}\ The authors thank W.~D.~Phillips and J.~V.~Porto for carefully reading the manuscript.
This work was partially supported by the National Institute of Standards and Technology, and the National Science Foundation through the Physics Frontier Center at the Joint Quantum Institute (PHY-1430094) and the Quantum Leap Challenge Institute for Robust Quantum Simulation (OMA-2120757).

\noindent {\bf Author contributions}\ E.A. and I.B.S. conceived of and developed the research direction.
E.A. performed and optimized the experiments, and developed the analysis framework. 
Theoretical aspects were led by I.B.S.
Both authors participated in the discussions of the results and contributed to the writing of this manuscript. 

\noindent {\bf Competing Interests}\ The authors declare no competing interests.

\noindent {\bf Code availability}\ The code used for analysis during the current study is available from the corresponding author on reasonable request.

\noindent {\bf Data availability}\ The datasets generated during and/or analyzed during the current study are available from the corresponding author on reasonable request.

\noindent {\bf Correspondence}\ Correspondence and requests for materials should be addressed to Emine~Altunta\c{s} (altuntas@ou.edu) and I.~B.~Spielman (ian.spielman@nist.gov).

%%%%%%%%%%%%%%%%%%%%%%%%%
\bibliography{main}

\begin{thebibliography}{42}%
\makeatletter
\providecommand \@ifxundefined [1]{%
 \@ifx{#1\undefined}
}%
\providecommand \@ifnum [1]{%
 \ifnum #1\expandafter \@firstoftwo
 \else \expandafter \@secondoftwo
 \fi
}%
\providecommand \@ifx [1]{%
 \ifx #1\expandafter \@firstoftwo
 \else \expandafter \@secondoftwo
 \fi
}%
\providecommand \natexlab [1]{#1}%
\providecommand \enquote  [1]{``#1''}%
\providecommand \bibnamefont  [1]{#1}%
\providecommand \bibfnamefont [1]{#1}%
\providecommand \citenamefont [1]{#1}%
\providecommand \href@noop [0]{\@secondoftwo}%
\providecommand \href [0]{\begingroup \@sanitize@url \@href}%
\providecommand \@href[1]{\@@startlink{#1}\@@href}%
\providecommand \@@href[1]{\endgroup#1\@@endlink}%
\providecommand \@sanitize@url [0]{\catcode `\\12\catcode `\$12\catcode
  `\&12\catcode `\#12\catcode `\^12\catcode `\_12\catcode `\%12\relax}%
\providecommand \@@startlink[1]{}%
\providecommand \@@endlink[0]{}%
\providecommand \url  [0]{\begingroup\@sanitize@url \@url }%
\providecommand \@url [1]{\endgroup\@href {#1}{\urlprefix }}%
\providecommand \urlprefix  [0]{URL }%
\providecommand \Eprint [0]{\href }%
\providecommand \doibase [0]{https://doi.org/}%
\providecommand \selectlanguage [0]{\@gobble}%
\providecommand \bibinfo  [0]{\@secondoftwo}%
\providecommand \bibfield  [0]{\@secondoftwo}%
\providecommand \translation [1]{[#1]}%
\providecommand \BibitemOpen [0]{}%
\providecommand \bibitemStop [0]{}%
\providecommand \bibitemNoStop [0]{.\EOS\space}%
\providecommand \EOS [0]{\spacefactor3000\relax}%
\providecommand \BibitemShut  [1]{\csname bibitem#1\endcsname}%
\let\auto@bib@innerbib\@empty
%</preamble>
\bibitem [{\citenamefont {von Neumann}(2018)}]{Neumann2018}%
  \BibitemOpen
  \bibfield  {author} {\bibinfo {author} {\bibfnamefont {J.}~\bibnamefont {von
  Neumann}},\ }\href@noop {} {\emph {\bibinfo {title} {Mathematical Foundations
  of Quantum Mechanics: New Edition}}}\ (\bibinfo  {publisher} {Princeton
  University Press},\ \bibinfo {address} {Princeton},\ \bibinfo {year}
  {2018})\BibitemShut {NoStop}%
\bibitem [{\citenamefont {Rossi}\ \emph {et~al.}(2018)\citenamefont {Rossi},
  \citenamefont {Mason}, \citenamefont {Chen}, \citenamefont {Tsaturyan},\ and\
  \citenamefont {Schliesser}}]{Rossi2018}%
  \BibitemOpen
  \bibfield  {author} {\bibinfo {author} {\bibfnamefont {M.}~\bibnamefont
  {Rossi}}, \bibinfo {author} {\bibfnamefont {D.}~\bibnamefont {Mason}},
  \bibinfo {author} {\bibfnamefont {J.}~\bibnamefont {Chen}}, \bibinfo {author}
  {\bibfnamefont {Y.}~\bibnamefont {Tsaturyan}},\ and\ \bibinfo {author}
  {\bibfnamefont {A.}~\bibnamefont {Schliesser}},\ }\bibfield  {title}
  {\bibinfo {title} {Measurement-based quantum control of mechanical motion},\
  }\href@noop {} {\bibfield  {journal} {\bibinfo  {journal} {Nature}\ }\textbf
  {\bibinfo {volume} {563}},\ \bibinfo {pages} {53} (\bibinfo {year}
  {2018})}\BibitemShut {NoStop}%
\bibitem [{\citenamefont {Terhal}(2015)}]{Terhal2015}%
  \BibitemOpen
  \bibfield  {author} {\bibinfo {author} {\bibfnamefont {B.~M.}\ \bibnamefont
  {Terhal}},\ }\bibfield  {title} {\bibinfo {title} {Quantum error correction
  for quantum memories},\ }\href@noop {} {\bibfield  {journal} {\bibinfo
  {journal} {Rev. Mod. Phys.}\ }\textbf {\bibinfo {volume} {87}},\ \bibinfo
  {pages} {307} (\bibinfo {year} {2015})}\BibitemShut {NoStop}%
\bibitem [{\citenamefont {Hosten}\ and\ \citenamefont
  {Kwiat}(2008)}]{Hosten2008}%
  \BibitemOpen
  \bibfield  {author} {\bibinfo {author} {\bibfnamefont {O.}~\bibnamefont
  {Hosten}}\ and\ \bibinfo {author} {\bibfnamefont {P.}~\bibnamefont {Kwiat}},\
  }\bibfield  {title} {\bibinfo {title} {Observation of the spin {H}all effect
  of light via weak measurements},\ }\href@noop {} {\bibfield  {journal}
  {\bibinfo  {journal} {Science}\ }\textbf {\bibinfo {volume} {319}},\ \bibinfo
  {pages} {787} (\bibinfo {year} {2008})}\BibitemShut {NoStop}%
\bibitem [{\citenamefont {Zhang}\ \emph {et~al.}(2015)\citenamefont {Zhang},
  \citenamefont {Datta},\ and\ \citenamefont {Walmsley}}]{Zhang2015}%
  \BibitemOpen
  \bibfield  {author} {\bibinfo {author} {\bibfnamefont {L.}~\bibnamefont
  {Zhang}}, \bibinfo {author} {\bibfnamefont {A.}~\bibnamefont {Datta}},\ and\
  \bibinfo {author} {\bibfnamefont {I.~A.}\ \bibnamefont {Walmsley}},\
  }\bibfield  {title} {\bibinfo {title} {Precision metrology using weak
  measurements},\ }\href@noop {} {\bibfield  {journal} {\bibinfo  {journal}
  {Phys. Rev. Lett.}\ }\textbf {\bibinfo {volume} {114}},\ \bibinfo {pages}
  {210801} (\bibinfo {year} {2015})}\BibitemShut {NoStop}%
\bibitem [{\citenamefont {Greve}\ \emph {et~al.}(2022)\citenamefont {Greve},
  \citenamefont {Luo}, \citenamefont {Wu},\ and\ \citenamefont
  {Thompson}}]{Greve2022}%
  \BibitemOpen
  \bibfield  {author} {\bibinfo {author} {\bibfnamefont {G.~P.}\ \bibnamefont
  {Greve}}, \bibinfo {author} {\bibfnamefont {C.}~\bibnamefont {Luo}}, \bibinfo
  {author} {\bibfnamefont {B.}~\bibnamefont {Wu}},\ and\ \bibinfo {author}
  {\bibfnamefont {J.~K.}\ \bibnamefont {Thompson}},\ }\bibfield  {title}
  {\bibinfo {title} {Entanglement-enhanced matter-wave interferometry in a
  high-finesse cavity},\ }\href@noop {} {\bibfield  {journal} {\bibinfo
  {journal} {Nature}\ }\textbf {\bibinfo {volume} {610}},\ \bibinfo {pages}
  {472} (\bibinfo {year} {2022})}\BibitemShut {NoStop}%
\bibitem [{\citenamefont {Freilich}\ \emph {et~al.}(2010)\citenamefont
  {Freilich}, \citenamefont {Bianchi}, \citenamefont {Kaufman}, \citenamefont
  {Langin},\ and\ \citenamefont {Hall}}]{Freilich2010}%
  \BibitemOpen
  \bibfield  {author} {\bibinfo {author} {\bibfnamefont {D.~V.}\ \bibnamefont
  {Freilich}}, \bibinfo {author} {\bibfnamefont {D.~M.}\ \bibnamefont
  {Bianchi}}, \bibinfo {author} {\bibfnamefont {A.~M.}\ \bibnamefont
  {Kaufman}}, \bibinfo {author} {\bibfnamefont {T.~K.}\ \bibnamefont
  {Langin}},\ and\ \bibinfo {author} {\bibfnamefont {D.~S.}\ \bibnamefont
  {Hall}},\ }\bibfield  {title} {\bibinfo {title} {Real-time dynamics of single
  vortex lines and vortex dipoles in a {B}ose-{E}instein condensate},\ }\href
  {https://doi.org/10.1126/science.1191224} {\bibfield  {journal} {\bibinfo
  {journal} {Science}\ }\textbf {\bibinfo {volume} {329}},\ \bibinfo {pages}
  {1182} (\bibinfo {year} {2010})}\BibitemShut {NoStop}%
\bibitem [{\citenamefont {Zeiher}\ \emph {et~al.}(2021)\citenamefont {Zeiher},
  \citenamefont {Wolf}, \citenamefont {Isaacs}, \citenamefont {Kohler},\ and\
  \citenamefont {Stamper-Kurn}}]{Zeiher2021}%
  \BibitemOpen
  \bibfield  {author} {\bibinfo {author} {\bibfnamefont {J.}~\bibnamefont
  {Zeiher}}, \bibinfo {author} {\bibfnamefont {J.}~\bibnamefont {Wolf}},
  \bibinfo {author} {\bibfnamefont {J.~A.}\ \bibnamefont {Isaacs}}, \bibinfo
  {author} {\bibfnamefont {J.}~\bibnamefont {Kohler}},\ and\ \bibinfo {author}
  {\bibfnamefont {D.~M.}\ \bibnamefont {Stamper-Kurn}},\ }\bibfield  {title}
  {\bibinfo {title} {Tracking evaporative cooling of a mesoscopic atomic
  quantum gas in real time},\ }\href@noop {} {\bibfield  {journal} {\bibinfo
  {journal} {Phys. Rev. X}\ }\textbf {\bibinfo {volume} {11}},\ \bibinfo
  {pages} {041017} (\bibinfo {year} {2021})}\BibitemShut {NoStop}%
\bibitem [{\citenamefont {Kristensen}\ \emph {et~al.}(2017)\citenamefont
  {Kristensen}, \citenamefont {Gajdacz}, \citenamefont {Pedersen},
  \citenamefont {Klempt}, \citenamefont {Sherson}, \citenamefont {Arlt},\ and\
  \citenamefont {Hilliard}}]{Kristensen2017}%
  \BibitemOpen
  \bibfield  {author} {\bibinfo {author} {\bibfnamefont {M.~A.}\ \bibnamefont
  {Kristensen}}, \bibinfo {author} {\bibfnamefont {M.}~\bibnamefont {Gajdacz}},
  \bibinfo {author} {\bibfnamefont {P.~L.}\ \bibnamefont {Pedersen}}, \bibinfo
  {author} {\bibfnamefont {C.}~\bibnamefont {Klempt}}, \bibinfo {author}
  {\bibfnamefont {J.~F.}\ \bibnamefont {Sherson}}, \bibinfo {author}
  {\bibfnamefont {J.~J.}\ \bibnamefont {Arlt}},\ and\ \bibinfo {author}
  {\bibfnamefont {A.~J.}\ \bibnamefont {Hilliard}},\ }\bibfield  {title}
  {\bibinfo {title} {Sub-atom shot noise {F}araday imaging of ultracold atom
  clouds},\ }\href@noop {} {\bibfield  {journal} {\bibinfo  {journal} {Journal
  of Physics B: Atomic, Molecular and Optical Physics}\ }\textbf {\bibinfo
  {volume} {50}},\ \bibinfo {pages} {034004} (\bibinfo {year}
  {2017})}\BibitemShut {NoStop}%
\bibitem [{\citenamefont {Christensen}\ \emph {et~al.}(2021)\citenamefont
  {Christensen}, \citenamefont {Vibel}, \citenamefont {Hilliard}, \citenamefont
  {Kruk}, \citenamefont {Paw\l{}owski}, \citenamefont {Hryniuk}, \citenamefont
  {Rza{\.z}ewski}, \citenamefont {Kristensen},\ and\ \citenamefont
  {Arlt}}]{Christensen2021}%
  \BibitemOpen
  \bibfield  {author} {\bibinfo {author} {\bibfnamefont {M.~B.}\ \bibnamefont
  {Christensen}}, \bibinfo {author} {\bibfnamefont {T.}~\bibnamefont {Vibel}},
  \bibinfo {author} {\bibfnamefont {A.~J.}\ \bibnamefont {Hilliard}}, \bibinfo
  {author} {\bibfnamefont {M.~B.}\ \bibnamefont {Kruk}}, \bibinfo {author}
  {\bibfnamefont {K.}~\bibnamefont {Paw\l{}owski}}, \bibinfo {author}
  {\bibfnamefont {D.}~\bibnamefont {Hryniuk}}, \bibinfo {author} {\bibfnamefont
  {K.}~\bibnamefont {Rza{\.z}ewski}}, \bibinfo {author} {\bibfnamefont {M.~A.}\
  \bibnamefont {Kristensen}},\ and\ \bibinfo {author} {\bibfnamefont {J.~J.}\
  \bibnamefont {Arlt}},\ }\bibfield  {title} {\bibinfo {title} {Observation of
  microcanonical atom number fluctuations in a {B}ose-{E}instein condensate},\
  }\href@noop {} {\bibfield  {journal} {\bibinfo  {journal} {Phys. Rev. Lett.}\
  }\textbf {\bibinfo {volume} {126}},\ \bibinfo {pages} {153601} (\bibinfo
  {year} {2021})}\BibitemShut {NoStop}%
\bibitem [{\citenamefont {Bohnet}\ \emph {et~al.}(2014)\citenamefont {Bohnet},
  \citenamefont {Cox}, \citenamefont {Norcia}, \citenamefont {Weiner},
  \citenamefont {Chen},\ and\ \citenamefont {Thompson}}]{Bohnet2014}%
  \BibitemOpen
  \bibfield  {author} {\bibinfo {author} {\bibfnamefont {J.~G.}\ \bibnamefont
  {Bohnet}}, \bibinfo {author} {\bibfnamefont {K.~C.}\ \bibnamefont {Cox}},
  \bibinfo {author} {\bibfnamefont {M.~A.}\ \bibnamefont {Norcia}}, \bibinfo
  {author} {\bibfnamefont {J.~M.}\ \bibnamefont {Weiner}}, \bibinfo {author}
  {\bibfnamefont {Z.}~\bibnamefont {Chen}},\ and\ \bibinfo {author}
  {\bibfnamefont {J.~K.}\ \bibnamefont {Thompson}},\ }\bibfield  {title}
  {\bibinfo {title} {Reduced spin measurement back-action for a phase
  sensitivity ten times beyond the standard quantum limit},\ }\href@noop {}
  {\bibfield  {journal} {\bibinfo  {journal} {Nature Photonics}\ }\textbf
  {\bibinfo {volume} {8}},\ \bibinfo {pages} {731} (\bibinfo {year}
  {2014})}\BibitemShut {NoStop}%
\bibitem [{\citenamefont {Krivoglaz}(1969)}]{Krivoglaz1969}%
  \BibitemOpen
  \bibfield  {author} {\bibinfo {author} {\bibfnamefont {M.~A.}\ \bibnamefont
  {Krivoglaz}},\ }\href@noop {} {\emph {\bibinfo {title} {Theory of X-Ray and
  Thermal Neutron Scattering by Real Crystals}}}\ (\bibinfo  {publisher}
  {Springer New York, NY},\ \bibinfo {year} {1969})\BibitemShut {NoStop}%
\bibitem [{\citenamefont {Lovesey}(1984)}]{Lovesey_1984}%
  \BibitemOpen
  \bibfield  {author} {\bibinfo {author} {\bibfnamefont {S.~W.}\ \bibnamefont
  {Lovesey}},\ }\href@noop {} {\emph {\bibinfo {title} {Theory of neutron
  scattering from condensed matter}}}\ (\bibinfo  {publisher} {Clarendon
  Press},\ \bibinfo {address} {United Kingdom},\ \bibinfo {year}
  {1984})\BibitemShut {NoStop}%
\bibitem [{\citenamefont {Ozeri}\ \emph {et~al.}(2005)\citenamefont {Ozeri},
  \citenamefont {Katz}, \citenamefont {Steinhauer},\ and\ \citenamefont
  {Davidson}}]{Ozeri2005}%
  \BibitemOpen
  \bibfield  {author} {\bibinfo {author} {\bibfnamefont {R.}~\bibnamefont
  {Ozeri}}, \bibinfo {author} {\bibfnamefont {N.}~\bibnamefont {Katz}},
  \bibinfo {author} {\bibfnamefont {J.}~\bibnamefont {Steinhauer}},\ and\
  \bibinfo {author} {\bibfnamefont {N.}~\bibnamefont {Davidson}},\ }\bibfield
  {title} {\bibinfo {title} {Colloquium: Bulk bogoliubov excitations in a
  {B}ose-{E}instein condensate},\ }\href@noop {} {\bibfield  {journal}
  {\bibinfo  {journal} {Rev. Mod. Phys.}\ }\textbf {\bibinfo {volume} {77}},\
  \bibinfo {pages} {187} (\bibinfo {year} {2005})}\BibitemShut {NoStop}%
\bibitem [{\citenamefont {Meinel}\ \emph {et~al.}(2022)\citenamefont {Meinel},
  \citenamefont {Vorobyov}, \citenamefont {Wang}, \citenamefont {Yavkin},
  \citenamefont {Pfender}, \citenamefont {Sumiya}, \citenamefont {Onoda},
  \citenamefont {Isoya}, \citenamefont {Liu},\ and\ \citenamefont
  {Wrachtrup}}]{Meinel2022}%
  \BibitemOpen
  \bibfield  {author} {\bibinfo {author} {\bibfnamefont {J.}~\bibnamefont
  {Meinel}}, \bibinfo {author} {\bibfnamefont {V.}~\bibnamefont {Vorobyov}},
  \bibinfo {author} {\bibfnamefont {P.}~\bibnamefont {Wang}}, \bibinfo {author}
  {\bibfnamefont {B.}~\bibnamefont {Yavkin}}, \bibinfo {author} {\bibfnamefont
  {M.}~\bibnamefont {Pfender}}, \bibinfo {author} {\bibfnamefont
  {H.}~\bibnamefont {Sumiya}}, \bibinfo {author} {\bibfnamefont
  {S.}~\bibnamefont {Onoda}}, \bibinfo {author} {\bibfnamefont
  {J.}~\bibnamefont {Isoya}}, \bibinfo {author} {\bibfnamefont {R.-B.}\
  \bibnamefont {Liu}},\ and\ \bibinfo {author} {\bibfnamefont {J.}~\bibnamefont
  {Wrachtrup}},\ }\bibfield  {title} {\bibinfo {title} {Quantum nonlinear
  spectroscopy of single nuclear spins},\ }\href@noop {} {\bibfield  {journal}
  {\bibinfo  {journal} {Nature Communications}\ }\textbf {\bibinfo {volume}
  {13}},\ \bibinfo {pages} {5318} (\bibinfo {year} {2022})}\BibitemShut
  {NoStop}%
\bibitem [{\citenamefont {Andrews}\ \emph {et~al.}(1996)\citenamefont
  {Andrews}, \citenamefont {Mewes}, \citenamefont {van Druten}, \citenamefont
  {Durfee}, \citenamefont {Kurn},\ and\ \citenamefont
  {Ketterle}}]{Andrews1996}%
  \BibitemOpen
  \bibfield  {author} {\bibinfo {author} {\bibfnamefont {M.~R.}\ \bibnamefont
  {Andrews}}, \bibinfo {author} {\bibfnamefont {M.-O.}\ \bibnamefont {Mewes}},
  \bibinfo {author} {\bibfnamefont {N.~J.}\ \bibnamefont {van Druten}},
  \bibinfo {author} {\bibfnamefont {D.~S.}\ \bibnamefont {Durfee}}, \bibinfo
  {author} {\bibfnamefont {D.~M.}\ \bibnamefont {Kurn}},\ and\ \bibinfo
  {author} {\bibfnamefont {W.}~\bibnamefont {Ketterle}},\ }\bibfield  {title}
  {\bibinfo {title} {Direct, nondestructive observation of a {B}ose
  condensate},\ }\href@noop {} {\bibfield  {journal} {\bibinfo  {journal}
  {Science}\ }\textbf {\bibinfo {volume} {273}},\ \bibinfo {pages} {84}
  (\bibinfo {year} {1996})}\BibitemShut {NoStop}%
\bibitem [{\citenamefont {Ketterle}(1999)}]{Ketterle1999}%
  \BibitemOpen
  \bibfield  {author} {\bibinfo {author} {\bibfnamefont {W.}~\bibnamefont
  {Ketterle}},\ }\bibinfo {title} {{B}ose-{E}instein condensation in atomic
  gases, proceedings of the international school of physics ``{E}nrico
  {F}ermi", course {CXL}}\ (\bibinfo  {publisher} {IOS Press},\ \bibinfo {year}
  {1999})\ Chap.\ \bibinfo {chapter} {Making, probing and understanding
  {B}ose-{E}instein condensates}, pp.\ \bibinfo {pages} {67--176}\BibitemShut
  {NoStop}%
\bibitem [{\citenamefont {Gajdacz}\ \emph {et~al.}(2013)\citenamefont
  {Gajdacz}, \citenamefont {Pedersen}, \citenamefont {M{\o}rch}, \citenamefont
  {Hilliard}, \citenamefont {Arlt},\ and\ \citenamefont
  {Sherson}}]{Gajdacz2013}%
  \BibitemOpen
  \bibfield  {author} {\bibinfo {author} {\bibfnamefont {M.}~\bibnamefont
  {Gajdacz}}, \bibinfo {author} {\bibfnamefont {P.~L.}\ \bibnamefont
  {Pedersen}}, \bibinfo {author} {\bibfnamefont {T.}~\bibnamefont {M{\o}rch}},
  \bibinfo {author} {\bibfnamefont {A.~J.}\ \bibnamefont {Hilliard}}, \bibinfo
  {author} {\bibfnamefont {J.}~\bibnamefont {Arlt}},\ and\ \bibinfo {author}
  {\bibfnamefont {J.~F.}\ \bibnamefont {Sherson}},\ }\bibfield  {title}
  {\bibinfo {title} {Non-destructive {F}araday imaging of dynamically
  controlled ultracold atoms},\ }\href@noop {} {\bibfield  {journal} {\bibinfo
  {journal} {Review of Scientific Instruments}\ }\textbf {\bibinfo {volume}
  {84}},\ \bibinfo {pages} {83105} (\bibinfo {year} {2013})}\BibitemShut
  {NoStop}%
\bibitem [{\citenamefont {Plenio}\ and\ \citenamefont
  {Knight}(1998)}]{Plenio1998}%
  \BibitemOpen
  \bibfield  {author} {\bibinfo {author} {\bibfnamefont {M.~B.}\ \bibnamefont
  {Plenio}}\ and\ \bibinfo {author} {\bibfnamefont {P.~L.}\ \bibnamefont
  {Knight}},\ }\bibfield  {title} {\bibinfo {title} {The quantum-jump approach
  to dissipative dynamics in quantum optics},\ }\href@noop {} {\bibfield
  {journal} {\bibinfo  {journal} {Rev. Mod. Phys.}\ }\textbf {\bibinfo {volume}
  {70}},\ \bibinfo {pages} {101} (\bibinfo {year} {1998})}\BibitemShut
  {NoStop}%
\bibitem [{\citenamefont {Wiseman}\ and\ \citenamefont
  {Milburn}(2011)}]{Wiseman2011}%
  \BibitemOpen
  \bibfield  {author} {\bibinfo {author} {\bibfnamefont {H.~M.}\ \bibnamefont
  {Wiseman}}\ and\ \bibinfo {author} {\bibfnamefont {G.~J.}\ \bibnamefont
  {Milburn}},\ }\href@noop {} {\emph {\bibinfo {title} {Quantum Measurement and
  Control}}}\ (\bibinfo  {publisher} {Cambridge University Press},\ \bibinfo
  {year} {2011})\BibitemShut {NoStop}%
\bibitem [{Kra()}]{KrausFootnote}%
  \BibitemOpen
  \href@noop {} {}\bibinfo {note} {This formalism is only valid for
  $\varphi\ll1$; a strong measurement can be modeled as the concatenation of
  many weak measurements.}\BibitemShut {Stop}%
\bibitem [{\citenamefont {Aharonov}\ \emph {et~al.}(1988)\citenamefont
  {Aharonov}, \citenamefont {Albert},\ and\ \citenamefont
  {Vaidman}}]{Aharonov1988}%
  \BibitemOpen
  \bibfield  {author} {\bibinfo {author} {\bibfnamefont {Y.}~\bibnamefont
  {Aharonov}}, \bibinfo {author} {\bibfnamefont {D.~Z.}\ \bibnamefont
  {Albert}},\ and\ \bibinfo {author} {\bibfnamefont {L.}~\bibnamefont
  {Vaidman}},\ }\bibfield  {title} {\bibinfo {title} {How the result of a
  measurement of a component of the spin of a spin-1/2 particle can turn out to
  be 100},\ }\href@noop {} {\bibfield  {journal} {\bibinfo  {journal} {Phys.
  Rev. Lett.}\ }\textbf {\bibinfo {volume} {60}},\ \bibinfo {pages} {1351}
  (\bibinfo {year} {1988})}\BibitemShut {NoStop}%
\bibitem [{Note1()}]{Note1}%
  \BibitemOpen
  \bibinfo {note} {To mitigate parasitic excitations that occur while the probe
  is applied, each measurement consisted of a pair of pulses of duration
  $t_m/2$ separated by a well chosen small delay.\cite
  {Altuntas2023a}}\BibitemShut {NoStop}%
\bibitem [{\citenamefont {Zernike}(1942)}]{Zernike1942}%
  \BibitemOpen
  \bibfield  {author} {\bibinfo {author} {\bibfnamefont {F.}~\bibnamefont
  {Zernike}},\ }\bibfield  {title} {\bibinfo {title} {Phase contrast, a new
  method for the microscopic observation of transparent objects},\ }\href@noop
  {} {\bibfield  {journal} {\bibinfo  {journal} {Physica}\ }\textbf {\bibinfo
  {volume} {9}},\ \bibinfo {pages} {686} (\bibinfo {year} {1942})}\BibitemShut
  {NoStop}%
\bibitem [{\citenamefont {Altunta{\c s}}\ and\ \citenamefont
  {Spielman}(2023{\natexlab{a}})}]{Altuntas2023a}%
  \BibitemOpen
  \bibfield  {author} {\bibinfo {author} {\bibfnamefont {E.}~\bibnamefont
  {Altunta{\c s}}}\ and\ \bibinfo {author} {\bibfnamefont {I.~B.}\ \bibnamefont
  {Spielman}},\ }\bibfield  {title} {\bibinfo {title} {Quantum back-action
  limits in dispersively measured {B}ose-{E}instein condensates},\ }\href
  {https://doi.org/10.1038/s42005-023-01181-5} {\bibfield  {journal} {\bibinfo
  {journal} {Communications Physics}\ }\textbf {\bibinfo {volume} {6}},\
  \bibinfo {pages} {66} (\bibinfo {year} {2023}{\natexlab{a}})}\BibitemShut
  {NoStop}%
\bibitem [{\citenamefont {Altunta{\c s}}\ and\ \citenamefont
  {Spielman}(2023{\natexlab{b}})}]{Altuntas2023}%
  \BibitemOpen
  \bibfield  {author} {\bibinfo {author} {\bibfnamefont {E.}~\bibnamefont
  {Altunta{\c s}}}\ and\ \bibinfo {author} {\bibfnamefont {I.~B.}\ \bibnamefont
  {Spielman}},\ }\bibfield  {title} {\bibinfo {title} {Weak-measurement-induced
  heating in {B}ose-{E}instein condensates},\ }\href
  {https://doi.org/10.1103/PhysRevResearch.5.023185} {\bibfield  {journal}
  {\bibinfo  {journal} {Phys. Rev. Res.}\ }\textbf {\bibinfo {volume} {5}},\
  \bibinfo {pages} {023185} (\bibinfo {year} {2023}{\natexlab{b}})}\BibitemShut
  {NoStop}%
\bibitem [{Note2()}]{Note2}%
  \BibitemOpen
  \bibinfo {note} {As discussed in the methods, the largest resolution-limited
  measurement strength in our experiment is $\varphi \approx 0.1$, at
  $g=1.0$.}\BibitemShut {Stop}%
\bibitem [{\citenamefont {Bogoliubov}(1947)}]{Bogoliubov1947}%
  \BibitemOpen
  \bibfield  {author} {\bibinfo {author} {\bibfnamefont {N.~N.}\ \bibnamefont
  {Bogoliubov}},\ }\bibfield  {title} {\bibinfo {title} {On the theory of
  superfluidity},\ }\href@noop {} {\bibfield  {journal} {\bibinfo  {journal}
  {J. Phys. (USSR)}\ }\textbf {\bibinfo {volume} {11}},\ \bibinfo {pages} {23}
  (\bibinfo {year} {1947})}\BibitemShut {NoStop}%
\bibitem [{MuF()}]{MuFootnote}%
  \BibitemOpen
  \href@noop {} {}\bibinfo {note} {The maximum 1D speed of sound
  $c=\sqrt{\mu/(2m)}$ was computed using the chemical potential $\mu =
  h\times0.8(1)\ {\rm kHz}$ measured for this dataset.}\BibitemShut {Stop}%
\bibitem [{\citenamefont {Altunta\ifmmode\mbox{\c{s}}\else\c{s}\fi{}}\ and\
  \citenamefont {Spielman}(2021)}]{Altuntas2021}%
  \BibitemOpen
  \bibfield  {author} {\bibinfo {author} {\bibfnamefont {E.}~\bibnamefont
  {Altunta\ifmmode\mbox{\c{s}}\else\c{s}\fi{}}}\ and\ \bibinfo {author}
  {\bibfnamefont {I.~B.}\ \bibnamefont {Spielman}},\ }\bibfield  {title}
  {\bibinfo {title} {Self-{B}ayesian aberration removal via constraints for
  ultracold atom microscopy},\ }\href
  {https://doi.org/10.1103/PhysRevResearch.3.043087} {\bibfield  {journal}
  {\bibinfo  {journal} {Phys. Rev. Research}\ }\textbf {\bibinfo {volume}
  {3}},\ \bibinfo {pages} {043087} (\bibinfo {year} {2021})}\BibitemShut
  {NoStop}%
\bibitem [{\citenamefont {Dressel}\ \emph {et~al.}(2014)\citenamefont
  {Dressel}, \citenamefont {Malik}, \citenamefont {Miatto}, \citenamefont
  {Jordan},\ and\ \citenamefont {Boyd}}]{Dressel2014}%
  \BibitemOpen
  \bibfield  {author} {\bibinfo {author} {\bibfnamefont {J.}~\bibnamefont
  {Dressel}}, \bibinfo {author} {\bibfnamefont {M.}~\bibnamefont {Malik}},
  \bibinfo {author} {\bibfnamefont {F.~M.}\ \bibnamefont {Miatto}}, \bibinfo
  {author} {\bibfnamefont {A.~N.}\ \bibnamefont {Jordan}},\ and\ \bibinfo
  {author} {\bibfnamefont {R.~W.}\ \bibnamefont {Boyd}},\ }\bibfield  {title}
  {\bibinfo {title} {Colloquium: Understanding quantum weak values: Basics and
  applications},\ }\href@noop {} {\bibfield  {journal} {\bibinfo  {journal}
  {Rev. Mod. Phys.}\ }\textbf {\bibinfo {volume} {86}},\ \bibinfo {pages} {307}
  (\bibinfo {year} {2014})}\BibitemShut {NoStop}%
\bibitem [{\citenamefont {Skinner}\ \emph {et~al.}(2019)\citenamefont
  {Skinner}, \citenamefont {Ruhman},\ and\ \citenamefont
  {Nahum}}]{Skinner2019}%
  \BibitemOpen
  \bibfield  {author} {\bibinfo {author} {\bibfnamefont {B.}~\bibnamefont
  {Skinner}}, \bibinfo {author} {\bibfnamefont {J.}~\bibnamefont {Ruhman}},\
  and\ \bibinfo {author} {\bibfnamefont {A.}~\bibnamefont {Nahum}},\ }\bibfield
   {title} {\bibinfo {title} {Measurement-induced phase transitions in the
  dynamics of entanglement},\ }\href@noop {} {\bibfield  {journal} {\bibinfo
  {journal} {Phys. Rev. X}\ }\textbf {\bibinfo {volume} {9}},\ \bibinfo {pages}
  {031009} (\bibinfo {year} {2019})}\BibitemShut {NoStop}%
\bibitem [{\citenamefont {Gullans}\ and\ \citenamefont
  {Huse}(2020)}]{Gullans2020}%
  \BibitemOpen
  \bibfield  {author} {\bibinfo {author} {\bibfnamefont {M.~J.}\ \bibnamefont
  {Gullans}}\ and\ \bibinfo {author} {\bibfnamefont {D.~A.}\ \bibnamefont
  {Huse}},\ }\bibfield  {title} {\bibinfo {title} {Dynamical purification phase
  transition induced by quantum measurements},\ }\href
  {https://doi.org/10.1103/PhysRevX.10.041020} {\bibfield  {journal} {\bibinfo
  {journal} {Phys. Rev. X}\ }\textbf {\bibinfo {volume} {10}},\ \bibinfo
  {pages} {041020} (\bibinfo {year} {2020})}\BibitemShut {NoStop}%
\bibitem [{\citenamefont {Noel}\ \emph {et~al.}(2022)\citenamefont {Noel},
  \citenamefont {Niroula}, \citenamefont {Zhu}, \citenamefont {Risinger},
  \citenamefont {Egan}, \citenamefont {Biswas}, \citenamefont {Cetina},
  \citenamefont {Gorshkov}, \citenamefont {Gullans}, \citenamefont {Huse},\
  and\ \citenamefont {Monroe}}]{Noel2022}%
  \BibitemOpen
  \bibfield  {author} {\bibinfo {author} {\bibfnamefont {C.}~\bibnamefont
  {Noel}}, \bibinfo {author} {\bibfnamefont {P.}~\bibnamefont {Niroula}},
  \bibinfo {author} {\bibfnamefont {D.}~\bibnamefont {Zhu}}, \bibinfo {author}
  {\bibfnamefont {A.}~\bibnamefont {Risinger}}, \bibinfo {author}
  {\bibfnamefont {L.}~\bibnamefont {Egan}}, \bibinfo {author} {\bibfnamefont
  {D.}~\bibnamefont {Biswas}}, \bibinfo {author} {\bibfnamefont
  {M.}~\bibnamefont {Cetina}}, \bibinfo {author} {\bibfnamefont {A.~V.}\
  \bibnamefont {Gorshkov}}, \bibinfo {author} {\bibfnamefont {M.~J.}\
  \bibnamefont {Gullans}}, \bibinfo {author} {\bibfnamefont {D.~A.}\
  \bibnamefont {Huse}},\ and\ \bibinfo {author} {\bibfnamefont
  {C.}~\bibnamefont {Monroe}},\ }\bibfield  {title} {\bibinfo {title}
  {Measurement-induced quantum phases realized in a trapped-ion quantum
  computer},\ }\href@noop {} {\bibfield  {journal} {\bibinfo  {journal} {Nature
  Physics}\ }\textbf {\bibinfo {volume} {18}},\ \bibinfo {pages} {760}
  (\bibinfo {year} {2022})}\BibitemShut {NoStop}%
\bibitem [{\citenamefont {Koh}\ \emph {et~al.}(2023)\citenamefont {Koh},
  \citenamefont {Sun}, \citenamefont {Motta},\ and\ \citenamefont
  {Minnich}}]{Koh2023}%
  \BibitemOpen
  \bibfield  {author} {\bibinfo {author} {\bibfnamefont {J.~M.}\ \bibnamefont
  {Koh}}, \bibinfo {author} {\bibfnamefont {S.-N.}\ \bibnamefont {Sun}},
  \bibinfo {author} {\bibfnamefont {M.}~\bibnamefont {Motta}},\ and\ \bibinfo
  {author} {\bibfnamefont {A.~J.}\ \bibnamefont {Minnich}},\ }\bibfield
  {title} {\bibinfo {title} {Measurement-induced entanglement phase transition
  on a superconducting quantum processor with mid-circuit readout},\
  }\href@noop {} {\bibfield  {journal} {\bibinfo  {journal} {Nature Physics}\
  }\textbf {\bibinfo {volume} {19}},\ \bibinfo {pages} {1314} (\bibinfo {year}
  {2023})}\BibitemShut {NoStop}%
\bibitem [{\citenamefont {Hoke}\ \emph {et~al.}(2023)\citenamefont {Hoke},
  \citenamefont {Ippoliti}, \citenamefont {Rosenberg}, \citenamefont {Abanin}
  \emph {et~al.}}]{Hoke2023}%
  \BibitemOpen
  \bibfield  {author} {\bibinfo {author} {\bibfnamefont {J.~C.}\ \bibnamefont
  {Hoke}}, \bibinfo {author} {\bibfnamefont {M.}~\bibnamefont {Ippoliti}},
  \bibinfo {author} {\bibfnamefont {E.}~\bibnamefont {Rosenberg}}, \bibinfo
  {author} {\bibfnamefont {D.}~\bibnamefont {Abanin}}, \emph {et~al.} (\bibinfo
  {collaboration} {Google Quantum AI and Collaborators}),\ }\bibfield  {title}
  {\bibinfo {title} {Measurement-induced entanglement and teleportation on a
  noisy quantum processor},\ }\href@noop {} {\bibfield  {journal} {\bibinfo
  {journal} {Nature}\ }\textbf {\bibinfo {volume} {622}},\ \bibinfo {pages}
  {481} (\bibinfo {year} {2023})}\BibitemShut {NoStop}%
\bibitem [{\citenamefont {Szyniszewski}\ \emph {et~al.}(2020)\citenamefont
  {Szyniszewski}, \citenamefont {Romito},\ and\ \citenamefont
  {Schomerus}}]{Szyniszewski2020}%
  \BibitemOpen
  \bibfield  {author} {\bibinfo {author} {\bibfnamefont {M.}~\bibnamefont
  {Szyniszewski}}, \bibinfo {author} {\bibfnamefont {A.}~\bibnamefont
  {Romito}},\ and\ \bibinfo {author} {\bibfnamefont {H.}~\bibnamefont
  {Schomerus}},\ }\bibfield  {title} {\bibinfo {title} {Universality of
  entanglement transitions from stroboscopic to continuous measurements},\
  }\href@noop {} {\bibfield  {journal} {\bibinfo  {journal} {Phys. Rev. Lett.}\
  }\textbf {\bibinfo {volume} {125}},\ \bibinfo {pages} {210602} (\bibinfo
  {year} {2020})}\BibitemShut {NoStop}%
\bibitem [{\citenamefont {Young}\ \emph {et~al.}(2021)\citenamefont {Young},
  \citenamefont {Gorshkov},\ and\ \citenamefont {Spielman}}]{Young2021}%
  \BibitemOpen
  \bibfield  {author} {\bibinfo {author} {\bibfnamefont {J.~T.}\ \bibnamefont
  {Young}}, \bibinfo {author} {\bibfnamefont {A.~V.}\ \bibnamefont
  {Gorshkov}},\ and\ \bibinfo {author} {\bibfnamefont {I.~B.}\ \bibnamefont
  {Spielman}},\ }\bibfield  {title} {\bibinfo {title} {Feedback-stabilized
  dynamical steady states in the {B}ose-{H}ubbard model},\ }\href@noop {}
  {\bibfield  {journal} {\bibinfo  {journal} {Phys. Rev. Research}\ }\textbf
  {\bibinfo {volume} {3}},\ \bibinfo {pages} {043075} (\bibinfo {year}
  {2021})}\BibitemShut {NoStop}%
\bibitem [{\citenamefont {Schuckert}\ and\ \citenamefont
  {Knap}(2020)}]{Schuckert2020}%
  \BibitemOpen
  \bibfield  {author} {\bibinfo {author} {\bibfnamefont {A.}~\bibnamefont
  {Schuckert}}\ and\ \bibinfo {author} {\bibfnamefont {M.}~\bibnamefont
  {Knap}},\ }\bibfield  {title} {\bibinfo {title} {Probing eigenstate
  thermalization in quantum simulators via fluctuation-dissipation relations},\
  }\href@noop {} {\bibfield  {journal} {\bibinfo  {journal} {Phys. Rev. Res.}\
  }\textbf {\bibinfo {volume} {2}},\ \bibinfo {pages} {043315} (\bibinfo {year}
  {2020})}\BibitemShut {NoStop}%
\bibitem [{\citenamefont {Kamenev}(2005)}]{kamenev2005}%
  \BibitemOpen
  \bibfield  {author} {\bibinfo {author} {\bibfnamefont {A.}~\bibnamefont
  {Kamenev}},\ }\href {https://arxiv.org/abs/cond-mat/0412296} {\bibinfo
  {title} {Many-body theory of non-equilibrium systems}} (\bibinfo {year}
  {2005}),\ \Eprint {https://arxiv.org/abs/cond-mat/0412296}
  {arXiv:cond-mat/0412296 [cond-mat.dis-nn]} \BibitemShut {NoStop}%
\bibitem [{\citenamefont {Altunta{\c s}}\ and\ \citenamefont
  {Spielman}(2023{\natexlab{c}})}]{Altuntas2023b}%
  \BibitemOpen
  \bibfield  {author} {\bibinfo {author} {\bibfnamefont {E.}~\bibnamefont
  {Altunta{\c s}}}\ and\ \bibinfo {author} {\bibfnamefont {I.~B.}\ \bibnamefont
  {Spielman}},\ }\bibfield  {title} {\bibinfo {title} {Direct calibration of
  laser intensity via {R}amsey interferometry for cold atom imaging},\
  }\href@noop {} {\bibfield  {journal} {\bibinfo  {journal} {Opt. Express}\
  }\textbf {\bibinfo {volume} {31}},\ \bibinfo {pages} {17893} (\bibinfo {year}
  {2023}{\natexlab{c}})}\BibitemShut {NoStop}%
\bibitem [{\citenamefont {Altunta{\c s}}\ \emph {et~al.}(2025)\citenamefont
  {Altunta{\c s}}, \citenamefont {Lena}, \citenamefont {Flannigan},
  \citenamefont {Daley},\ and\ \citenamefont {Spielman}}]{Altuntas2025}%
  \BibitemOpen
  \bibfield  {author} {\bibinfo {author} {\bibfnamefont {E.}~\bibnamefont
  {Altunta{\c s}}}, \bibinfo {author} {\bibfnamefont {R.~G.}\ \bibnamefont
  {Lena}}, \bibinfo {author} {\bibfnamefont {S.}~\bibnamefont {Flannigan}},
  \bibinfo {author} {\bibfnamefont {A.~J.}\ \bibnamefont {Daley}},\ and\
  \bibinfo {author} {\bibfnamefont {I.~B.}\ \bibnamefont {Spielman}},\
  }\bibfield  {title} {\bibinfo {title} {Dynamical structure factor from weak
  measurements},\ }\href@noop {} {\bibfield  {journal} {\bibinfo  {journal}
  {Quantum Science and Technology}\ }\textbf {\bibinfo {volume} {10}},\
  \bibinfo {pages} {035045} (\bibinfo {year} {2025})}\BibitemShut {NoStop}%
\end{thebibliography}%
%%%%%%%%%%%%%%%%%%%%%%%%%

\newpage
% \onecolumn
% \appendix
%----------------------------------------------------------------------------------------
% %%%%%%%%%%%%%%%%%            SUPPLEMENTARY INFORMATION             %%%%%%%%%%%%%%%%%  %
% %---------------------------------------------------------------------------------------
\pagebreak
\widetext
\begin{center}
	\textbf{\large Supplementary Information for ``Repeated weak measurements: watching quantum correlations evolve''}
\end{center}
%%%%%%%%%% Merge with supplemental materials %%%%%%%%%%
%%%%%%%%%% Prefix a "S" to all equations, figures, tables and reset the counter %%%%%%%%%%
\setcounter{equation}{0}
\setcounter{figure}{0}
\setcounter{table}{0}
\makeatletter
\renewcommand{\theequation}{S\arabic{equation}}
\renewcommand{\thefigure}{S\arabic{figure}}
\renewcommand{\bibnumfmt}[1]{[S#1]}
\renewcommand{\citenumfont}[1]{S#1}
% \renewcommand{\thepage}{S\arabic{page}}  

%----------------------------------------------------------------------------------------
%%%%%%%%%%%%%%%%%    App: conventional parameters  %%%%%%%%%%%%%%%%%
%----------------------------------------------------------------------------------------
% \section*{Methods}

\section{Fourier transform convention} \label{app:fourier}
\ 
There are many sensible options for discrete Fourier transform pairs; since our analysis code is internally dimensionless we take the symmetric text-book 1D pair
\begin{align*}
f_x &= \frac{1}{\sqrt{N_x}} \sum_k \tilde f_k e^{ikx} &&\leftrightarrow&
\tilde f_k &=  \frac{1}{\sqrt{N_x}} \sum_x f_x e^{-ikx},
\end{align*}
with the ``physics'' convention for wavenumber, and where the tilde marks Fourier transformed quantities.
It is common in the optics literature to use $2\pi k x$ rather than $kx$ in Fourier transforms; to make connection with this literature, the wavevector axes in Figs.~2{\bf c},~\ref{FigSM:PSDsteps} and \ref{FigSM:PCA_NextShot} are scaled by $2\pi$.
In two-dimensions (2D) we use the obvious extension with $\sqrt{N_x}$ replaced with $\sqrt{N_x N_y}$, and pixel coordinates ${\bf r} = (x, y)$.

\section{Correlation and cross-correlation analysis}\label{app:analysis}
\ 
Here we detail our cross-correlation analysis procedure.
For every set of experimental parameters (delay time, temperature, and etc.) we create an ensemble by repeating the experiment $M=128$ times, and label each repetition by $j$.
Four in-situ PCI images are recorded in each image, spaced by the same time-delay.
We begin by describing our standard process for obtaining the number fluctuations $\delta n^{(j)}_{\bf r}$ associated with each member of this ensemble.

\begin{enumerate}

\item {\it Raw PCI signal} Each PCI measurement consists of the three raw images: with-atoms $I^{(j)}_{+,{\bf r}}$, probe only $I^{(j)}_{0,{\bf r}}$ and dark $I^{(j)}_{D,{\bf r}}$. 
These data were taken with an Andor DU-888UU3 EMCCD with an $1024\times1024$ array of $13\ \mu{\rm m}$ pixels, each reduced to an object-plane size of $\approx 0.36\ \mu{\rm m}$ by imaging system's $\times36$ magnification.
This is far smaller than the $\approx2\ \mu{\rm m}$ resolution of our imaging system.
In order to eliminate any spurious background illumination, or dark counts, we subtract 
the ensemble averaged dark frame $\overline{I^{(j)}_D}$ from $I^{(j)}_+$ and $I^{(j)}_-$.  
For simplicity's sake, in what follows we make the replacement $I^{(j)}_\pm - \overline{I^{(j)}_D} \rightarrow I^{(j)}_\pm$, and omit the position index ${\bf r}$ in unambiguous cases.
The PCI signal is given by
\begin{align}
g_{\rm PCI} = 1 - \frac{I^{(j)}_+ }{I^{(j)}_-}, \label{eq:PCI}
\end{align}
where we suppressed the positional dependence for clarity.
Figure~\ref{FigSM:PCIsteps}{\bf a} shows a single-shot PCI image.
 
\item {\it Initial noise and artifact reduction} We use the ensemble averaged dark frame $\overline{I^{(j)}_D}$ and employ principal component analysis (PCA) techniques\cite{Altuntas2023b} in real space on the ensemble of probe images $\{I^{(j)}_0\}_j$ to obtain a set of optimized probes $\{I^{(j)}_{\rm PCA}\}_j$.   
This has the effect of both reducing noise (readout noise is nearly eliminated in the averaged dark frames and PCA-reconstructed probes have reduced shot noise) and imaging artifacts (PCA-reconstructed probe images have fringe patterns nearly ideally matched to their with-atoms counterparts). 
Figure~\ref{FigSM:PCIsteps}{\bf b} shows the same data as in {\bf a} following this process.
Comparing {\bf a} and {\bf b} shows that this greatly reduces the fringes (from etaloning in our glass cell) present in the initial image.
For reference, Fig.~\ref{FigSM:PCIsteps}{\bf c} shows data such as in {\bf b} averaged over 128 experimental repetitions.

\begin{figure}[htb]
\begin{center}
\includegraphics{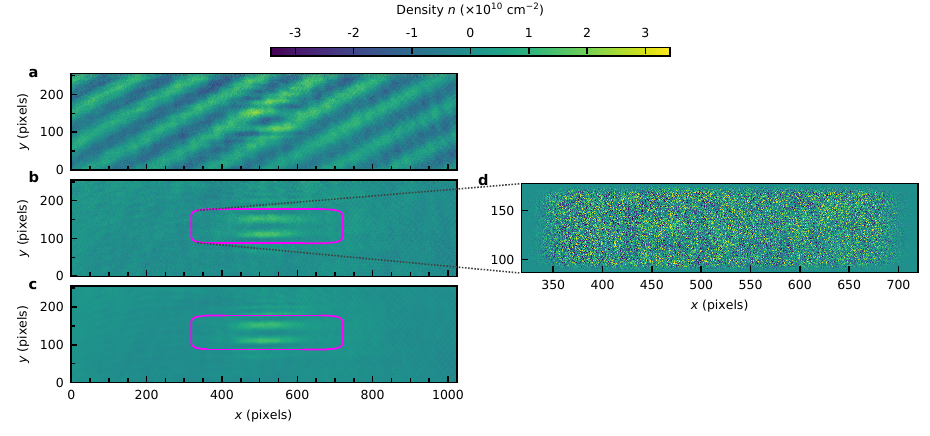}
\end{center}
\caption[Real space analysis steps]{
\textbf{Real space analysis steps.}
M2 images from the dataset employed in Fig.~\ref{Fig2:2FK_CCF} (with measurement strength $g'\approx 1$).
{\bf a} Raw PCI signal computed according to Eq.~\eqref{eq:PCI} showing the aberrated {\it in situ} atomic cloud along with interference fringes on many length scales.
{\bf b} PCI signal computed using an optimized probe reconstructed from a linear combination of 128 separate probe images with greatly reduced interference fringes.
{\bf c} PCI signal averaged over an ensemble of 128 measurements.
The magenta curve indicates the extent of the 2D squircle-shaped Tukey window applied prior to {\bf d}.
{\bf d} Zoomed in view of a single-shot $\delta n$ measurement (with contributions from atom-number fluctuations, photo-electron shot noise, and remnant technical artifacts) with the real space window applied. 
The $\delta n$ signal is amplified by $\times 5$ compared to data in {\bf a} to {\bf c} to increase visibility.
}
\label{FigSM:PCIsteps}
\end{figure}

\item {\it Atomic density} We compute the PCI signal using these images leading to the 2D column density $n^{(j)} = 2\delta / (\sigma_0 \Gamma)  g_{\rm PCI}$, where $\sigma_0$ is the resonant scattering cross-section. 

\item {\it Density fluctuations} One might be tempted to compute the density fluctuations as $\delta n^{(j)} = n^{(j)} - \overline{n}$, however, experimental shot-to-shot differences in quantities such as total atom number, trap center and so forth lead to a large background using this simple approach.
Instead we create a PCA basis on the full ensemble of images $\{n^{(j)}\}_j$ and reconstruct an optimal average $\overline{n_{{\rm PCA}}^{(j)}}$  for each image built from a linear combination of the most significant principle components (for an ensemble of size 128 we use 7 components).  
This leads to a background free map of fluctuations $\delta n^{(j)} = n^{(j)} - \overline{n_{{\rm PCA}}^{(j)}}$.

\item {\it Real space window} We reduce noise and artifacts by applying the squircle-shaped Tukey window with boundary plotted in magenta in Fig.~\ref{FigSM:PCIsteps}{\bf c}.
The horizontal extent of this window was selected to be $\approx 10\ \%$ larger than the atomic cloud inferred from time of flight (TOF) measurements. 
Because our imaging system is aberrated, the vertical extent is optimized to pass as much signal as possible while simultaneously minimizing the photon shot noise present in all regions of the electron multiplied charge coupled device (EMCCD). 

\end{enumerate}

The statistical properties of the number fluctuations can be parameterized by the power spectral density (PSD) usually defined as ${\rm PSD}_{\bf k}[\delta n^{(j)}_{\bf r}] \equiv {|\tilde {\delta n}_{\bf k}^{(j)}|^2}$.
The $k$-space circle defining the imaging system's numerical aperture (NA) also determines the region of the PSD (for $|{\bf k}| < k_{\rm NA}$) that can contain meaningful signal.
Here, $k_{\rm NA} = {\rm NA} \times k_0 $ with wavenumber $k_{0}$\,=\,$2\pi / \lambda$ for a laser of wavelength $\lambda$.
As discussed below, this simple procedure yields excess noise as well as artifacts.

\begin{figure}[htb]
\begin{center}
\includegraphics{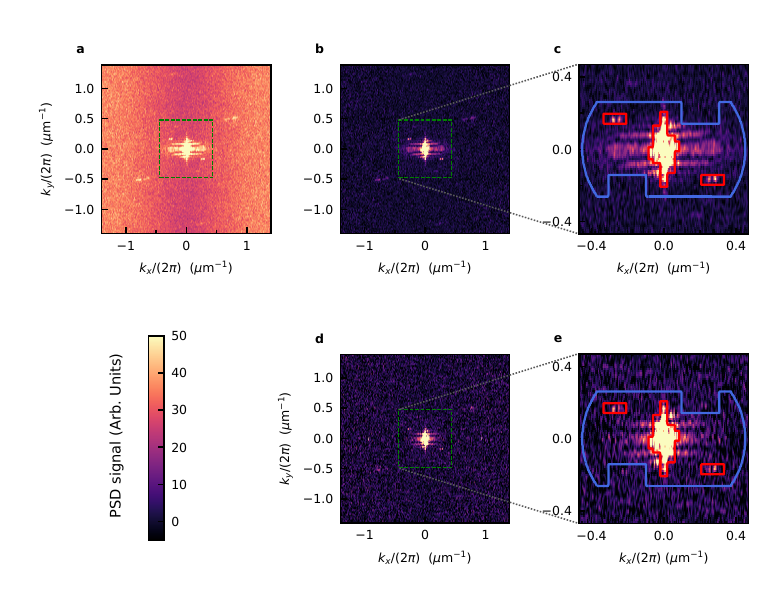}
\end{center}
\caption[2D correlation function analysis steps]{
\textbf{2D correlation function analysis steps.}
Applied to the same data as in Fig.~\ref{FigSM:PCIsteps}. 
(I) PSD analysis of M2 from this dataset (at $\delta t = 0.45$ ms ).
{\bf a} Ensemble averaged PSD. 
The green-dashed square marks the Fourier mask---fully enclosing the NA limited range of wavevectors---used in the computation of the photon shot noise background.
{\bf b} PSD with the photon shot noise background subtracted.
{\bf c} Expanded view of the PSD.
The blue curve outlines the full symmetrized aperture resulting from the NA-limit of our objective along with further regions occulted by our experimental geometry.
The red curves outline the small-$k$ masks applied to remove artifacts that survive our real space analysis.
(II) CPSD analysis at $\delta t = 0.45$ ms. 
Because photon-shot noise does not contribute a background to CPSD data, there is no partner figure to {\bf a}.
{\bf d} and {\bf e} Full-scale and zoomed CPSD data showing structure similar to the PSD data in {\bf b} and {\bf c}.
All data were averaged over an ensemble of 128 measurements.
}
\label{FigSM:PSDsteps}
\end{figure}

Figure~\ref{FigSM:PSDsteps}{\bf a} shows the PSD obtained from the same dataset as in Fig.~\ref{FigSM:PCIsteps}{\bf c}.  
Data outside the green-dashed rectangle is outside the NA-window, and any PSD signal outside the NA-window results either from sensor artifacts or photon shot noise.
Photon shot noise contributes both an overall offset (because it is spatially uncorrelated) and noise (``noise in the noise'') to the PSD signal.
Notice that the smooth background signal (from photon shot noise) is not constant.
This is an artifact of operating our EMCCD sensor in fast kinetics mode: in standard imaging modes the photon shot noise background is independent of $k$.

The gigantic signal for $|{\bf k}|/(2\pi)\lesssim 0.1\ \mu{\rm m}^{-1}$ results from imaging artifacts that were not fully eliminated by the analysis procedure described above. 
The largest contribution stems from remnant structure (interference fringes) from the probe beam, which we confirmed by following our analysis procedure on data with no atoms present.

Lastly, the zebra-like pattern in the PSD encodes imaging system aberrations.
While these can be compensated for with regularization techniques as in Ref.~\onlinecite{Altuntas2021}, we observed marginal improvement in our final data quality.
We thereby did not incorporate the regularization step in our analysis pipeline.

We now turn to our techniques for managing these artifacts to obtain final Fourier space data.

\begin{enumerate}

\item {\it Photon shot noise removal}  
For PSD analysis, we remove the photon shot noise offset by first masking the region of the PSD inside the NA-window, then averaging the remainder, and subtracting it from the original PSD\cite{Altuntas2021}.
This still leaves behind fluctuations away from zero (noise in the noise) outside the NA-window, that we therefore discard by setting it to zero.
The PSD signal in Fig~\ref{FigSM:PSDsteps}{\bf b} shows the result of this process.
In our measurements, the NA-window is further refined to account for components of our apparatus that occult part of the aperture, as described in Ref.~\onlinecite{Altuntas2021}.
The blue curve in Fig.~\ref{FigSM:PSDsteps}{\bf c} illustrates the final window function along with a magnified view of the PSD shown in Fig.~\ref{FigSM:PSDsteps}{\bf b}.

\item {\it Small-$k$ mask} 
We eliminate the small-$k$ artifacts present within the NA-window with a mask function empirically determined to fully shroud these spurious features.
This is shown by the red outline in Fig.~\ref{FigSM:PSDsteps}{\bf c}. 
(In the next section, we describe a more sophisticated PCA based analysis that yields improved noise reduction in the cross-correlation signal.)

\item {\it Cross-power spectral density (CPSD)} 
Following the complete analysis described above, we obtain the cross-power spectral density ${\rm CPSD}^{(j)}_{\bf k} \equiv [{\tilde {\delta n}^{(j)}_{\bf k}(t=0)] [\tilde {\delta n}^{(j)}_{\bf k}(t=\delta t)]^* }$ for every member of the ensemble.
Average CPSD from the dataset used above is shown in Fig.~\ref{FigSM:PSDsteps}{\bf d} with a zoom view in {\bf e}.
We note that photon-shot noise does not contribute an offset to CPSD data because the two images have uncorrelated photon shot noise realizations, allowing us to omit step 1 for CPSD data.
\end{enumerate}

Thus far in our analysis we have operated in the natural 2D space provided by our image data.
However, because our BECs are highly elongated along $\ex$ the long-wavelength phonon excitations correspond to waves traveling along $\ex$.  
We therefore now turn to our procedure for extracting 1D correlation and cross-correlations signals from the 2D CPSD.

First, recall that by the Fourier-convolution (Wiener–Khinchin) theorem, the cross correlation function (CCF)
\begin{align}
{\rm CCF}_{\delta {\bf r}} &= \sum_{\bf r} \delta n_{\bf r}(t=0) \delta n_{{\bf r} + \delta {\bf r}} (t=\delta t) = \frac{1}{\sqrt{N_x N_y}}\sum_{\bf k}  e^{i{\bf k}\cdot\delta{\bf r}} {\rm CPSD}_{\bf k}
\end{align}
is directly related to the CPSD.
There are therefore two natural approaches for computing 1D correlation functions from ${\rm CPSD}_{\bf k}$: one can either average ${\rm CPSD}_{\bf k}$ along $k_y$ to obtain a 1D ${\rm CPSD}_{k_x}$ and then take the inverse Fourier transform, or obtain ${\rm CCF}_{\delta {\bf r}}$ and then average along $y$.
Notice that the first process ``$k_y$-avaraging'' is equivalent to evaluating ${\rm CCF}_{x, y=0}$, and the second ``$y$-averaging'' is equivalent to evaluating ${\rm CPSD}_{k_x, k_y=0}$ and then computing the inverse Fourier transform up to a factor of $\sqrt{N_y}$.
Applying either technique to the CPSD signal yield a slightly different 1D CCF.
In our analysis we employ the ``$y$-averaging'' procedure as it focuses on those excitations traveling exclusively along the long axis of the BEC.

%%%%%%%%%%%%%%%%% %%%%%%%%%%%%%%%%% %%%%%%%%%%%%%%%%% %%%%%%%%%%%%%%%%% 
\section{PCA based small-$k$ artifact removal}\label{app:PCA_analysis}
\ 
Here we describe a PCA technique that suppresses small-$k$ CPSD artifacts without explicit masking, thereby making the naive small-$k$ mask approach described above redundant.

\begin{figure}[htb]
\begin{center}
\includegraphics{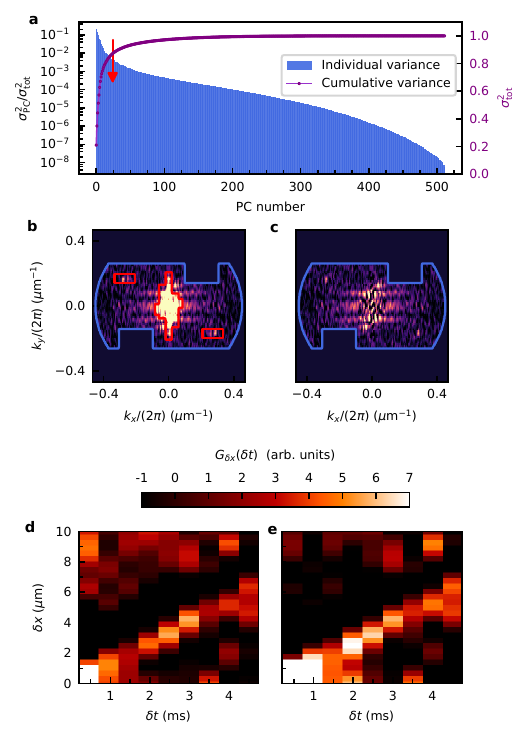}
\end{center}
\caption[Correlation function computation with the PCA based small-$k$ masking]{
\textbf{PCA based small-$k$ masking.}
Applied to the same data as in Fig.~\ref{FigSM:PCIsteps} that lead to the 1D CCFs in Fig.~\ref{Fig2:2FK_CCF}.
{\bf a} Individual (blue) and cumulative (purple) explained variance of the PCA basis. 
The red arrow marks the component number for which the cumulative variance reaches $\approx 0.87$.
{\bf b} Zoomed view of CPSD at $\delta t = 0.45~\rm{ms}$.
Red curves outline the small-$k$ mask applied in the cross-correlation analysis. 
{\bf c} CPSD with PCA based masking applied. 
{\bf d} Symmetrized Van Hove function obtained with the hard-wall small-$k$ masking illustrated in {\bf a}.
{\bf e} Symmetrized Van Hove function obtained using the PCA-based masking.
}
\label{FigSM:PCA_NextShot}
\end{figure}

We construct a basis of 2D CPSD built from $t=0$ and $t=\delta t$ images taken from different experimental realizations with the same parameters. 
As a result, stochastic noise (projection noise, thermal excitations, and technical noise) in the first measurement M1 is uncorrelated with that in the second measurement M2.
We find that these CPSDs still host the small-$k$ structure in the CPSD.
In principle, a dataset of $M$ measurements could generate up to $M^2 - M$ such CPSDs by combining M1 and M2 from every pairwise combination of distinct realizations.
Owing to computational memory limitations we limited our basis dimension to $512$.
Figure~\ref{FigSM:PCA_NextShot}{\bf a} plots the individual and cumulative explained variance of the orthonormal PCA basis, and as marked by the red arrow, we retain the $\approx 20$ most significant components.
We then subtract the contribution of each of these components from true single-shot CPSD data within the original mask regions (this avoids adding noise outside the region where artifacts are absent).
This process reduces the amplitude of the artifacts by about a factor of ten and their extent in $k$-space is vastly diminished.

Figure~\ref{FigSM:PCA_NextShot}{\bf(b-e)} shows the results of this PCA based analysis.
Panel {\bf b} shows the initial PCA signal with small-$k$ artifacts present, and the masked regions described above.
In comparison, {\bf c} shows that the PCA based analysis greatly reduces the small-$k$ artifacts. 
Figures~\ref{FigSM:PCA_NextShot}{\bf(d,e)} compare the final real-space correlation signals; the PCA-based technique yields a reduction in noise at larger values of $x$.

The CCF/QWV data in Figs. 3 and 4 show a persistent peak at $\delta x = 0$.
Operationally this results from a small positive offset in the Fourier transform that is not present in the dataset presented in Fig. 2.  
We have identified no causal origin for this artifact.
As such this feature was not included in any of the theoretical model functions used in Fig 2 and Fig 4.

%%%%%%%%%%%%%%%%% %%%%%%%%%%%%%%%%% %%%%%%%%%%%%%%%%% %%%%%%%%%%%%%%%%% 
\section{CCF line-shape model}\label{app:lineshape_model}
\
In this section, we elaborate on the CCF line-shape model yielding the blue dashed curves in Fig.~\ref{Fig2:2FK_CCF}{\bf a} and in subsequent fits.
This model predicts correlations between Bogoliubov excitations (under a local-density-approximation), viewed through our imperfect imaging system. 

Although to this point, we framed the creation and detection of excitations in the context real-space measurements, it can equivalently be described as a momentum-space scattering process\cite{Altuntas2023a}.
From this perspective, low momentum phonons are created by small-angle forward scattering of the incident probe laser.
For small imparted transverse momentum, i.e. $|k_x| < k_{\rm NA}$, these phonon excitations can be resolved by our imaging system and are the focus of our manuscript.
However, excitations with small transverse momentum can also be created by back-scattering processes that, in addition, impart a large forward momentum $\approx 2 k_R$, where $k_R$ is the single-photon recoil momentum.
For our chemical potential, the healing length is $\xi \approx 0.25\ \mu{\rm m}$, making these excitations particle-like (since $\xi^{-1} \approx 0.5 \times k_R$), well described by the free particle dispersion $\hbar^2 {\bf k}^2 / 2m$.
Thus optical scattering yields two distinct ``flavors'' of resolvable excitations: transverse-propagating phonons and forward-scattered free atoms.

As evidenced in our correlation analysis, the low energy phonon-excitations are slowly moving and remain relevant even to our longest time delay $\delta t$.
By contrast, forward-scattered atoms rapidly depart the BEC along its narrow transverse direction (with Thomas-Fermi radius of just $R_z \approx 3~\mu\rm{m}$), when their visibility rapidly declines.
As a result, excitations originating from small-angle backward scattering heavily contribute to the observed CCF at short time scales, and rapidly decay beyond the $\approx 0.5~{\rm ms}$ time scale, given the $5.9~\mu\rm{m/ms}$ single-photon recoil velocity. 

\begin{figure}[htb]
\begin{center}
\includegraphics{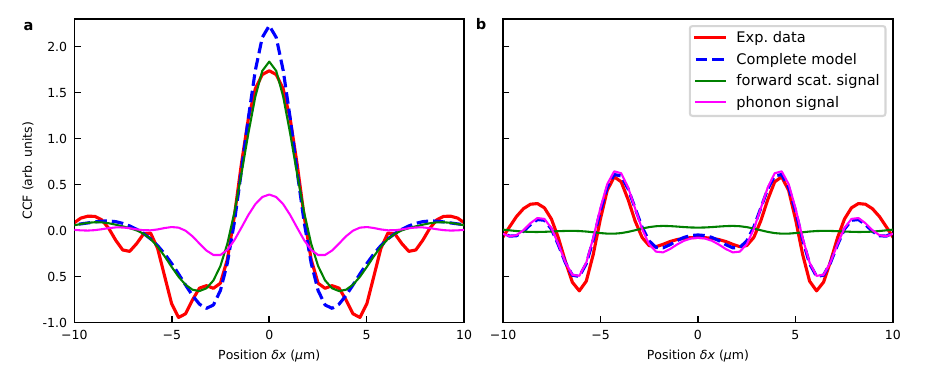}
\end{center}
\caption[CCF line-shape model contributions]{
{\bf CCF line-shape model.}
Separate contributions to the line-shape from different scattering processes at $t=0.45\ {\rm ms}$ and $t=3.0\ {\rm ms}$ are shown in {\bf a} and {\bf b} respectively.
In both panels, we plot the experimental CCF (solid red), the complete line-shape model (dashed blue), the forward scattered contribution (solid green) and the phonon contribution (solid magenta).
The experimental data and complete model are from Fig.~2{\bf a}.
}
\label{FigSM:ModelFit}
\end{figure}

The model function combines the contributions arising from low energy phonons and forward scattered atoms with amplitudes parametrized by $s_p$ and $s_f$ respectively.
These contributions, including the effect of imaging aberrations, are plotted along with their sum in Fig.~\ref{FigSM:ModelFit}, for the same values of $\delta t$ used in Fig.~2{\bf a}.
We model the disappearance of forward-scattered excitations with an exponential decay function $s_f = h_f \, e^{-\gamma_f \delta t}$, defined in terms of a zero-time amplitude $h_f$, the decay rate $\gamma_f$, and, as in the main manuscript, the time interval $\delta t$.
This is evidenced by the significant reduction in the forward scattered model curves from Fig.~\ref{FigSM:ModelFit}{\bf a} ($\delta t = 0.45\ {\rm ms}$) to {\bf b} ($\delta t = 3.0~\rm{ms}$).
By contrast the phonon amplitude $s_p$ is found to be independent of $\delta t$.

We perform fits to the experimental data using this model in two ways.
For example, for Van Hove data sets we either perform individual fits for each value of $\delta t$, or global fits to complete dataset.
In principle each CCF can be modeled with the parameters $s_p$, $h_f$, $\gamma_f$, as well as the speed of sound $c$. 
In the case of individual fitting, $c$ and $s_p$ vary independently for each $dt$.
On the other hand, for global fits the values $c$ and $s_p$ are shared across the entire dataset. 
For both global and individual fitting, $s_f$ is modeled by the exponential decay function given above with all fit parameters shared.
In Figs.~\ref{Fig2:2FK_CCF}{\bf b}, 4{\bf a} and  4{\bf b}, the individual fitting procedure yields the circles, whereas the global fit is used to compute the $c t$ line. 
We perform analogous individual and global fits to the $R_c$ dependent data shown in Fig.~4{\bf c}.

%%%%%%%%%%%%%%%%% %%%%%%%%%%%%%%%%% %%%%%%%%%%%%%%%%% %%%%%%%%%%%%%%%%% 
\section{Theoretical concept}\label{app:theory_concept}
\ 
Our physical apparatus sequentially generates individual elements of an ensemble consisting of $M$ BECs described by the density operator $\hat \rho = \sum_j\ket{\Psi^{(j)}}\bra{\Psi^{(j)}}/M$; each BEC, labeled by $j$, is an isolated quantum system with $t=0$ wavefunction $\ket{\Psi^{(j)}}$.
We denote the ensemble expectation value of a Hermitian operator $\hat O$ as
\begin{align}
\langle \hat O \rangle &= {\rm Tr} (\hat O \hat \rho ) = \frac{1}{M} \sum_j \langle \hat O \rangle^{(j)},
\end{align}
and add a superscript $(j)$ to denote the expectation value of individual ensemble members, i.e.,  $\langle \hat O \rangle^{(j)} \equiv \bra{\Psi^{(j)}}\hat O\ket{\Psi^{(j)}}$.

As experimentalists we have access to neither the single-instance wavefunctions nor even the density operator, and must instead focus on our actual observable of atomic number. 
In terms of the wave function, the $i$-th measurement outcome, taken at time $t_i$ (in the main manuscript, we had just two measurements, for which $t_1 = 0$ and $t_2 = \delta t$), for the $j$-th member of our ensemble would be
\begin{align}
n^{(j)}_{x,i} &= \langle \hat n_{x} \rangle^{(j)\prime}_i + \frac{m^{(j)}_{x,i}}{\varphi}
\end{align}
in terms of $\langle \hat n_{x} \rangle^{(j)\prime}_i = \bra{\Psi^{(j)\prime}_i} \hat n_x \ket{\Psi^{(j)\prime}_i}=\bra{\Psi^{(j)\prime}(t_i)} \hat n_x \ket{\Psi^{(j)\prime}(t_i)}$, the measurement strength $\varphi$ (see Supplementary Information Sect.~\ref{app:strength} for the dependence of $\varphi$ on physical parameters), and the zero-mean Gaussian random variable $m^{(j)}_{x,i}$ with $\overline{m^{(j_1)}_{x_1,i_1} m^{(j_2)}_{x_2,i_2}} = \delta_{j_1,j_2} \delta_{i_1,i_2} \delta_{x_1,x_2} / 2$.
Primed quantities reference the true state at time $t_i$ including the backaction from all prior measurements.
Quantities without primes include only unitary evolution (as if no measurements had taken place).
Averaged over the experimental ensemble this reduces to the observable $\overline{n^{(j)}_{x,i}} = \langle \hat n_{x} \rangle^\prime_i$.
The noise
\begin{align}
\delta n^{(j)}_{x,i} &\equiv \left[\langle \hat n_{x} \rangle^{(j)\prime}_i - \langle \hat n_{x} \rangle^\prime_i\right] + \frac{m^{(j)}_{x,i}}{\varphi}
\end{align}
contains additional information: the first term describes statistical noise and the second quantifies the quantum projection noise.

As usual, the backaction associated with each measurement is quantified by a Kraus operator\cite{Wiseman2011}, with lowest order expansion
\begin{align}
\hat K^{(j)}_i &\approx 1 + \varphi \sum_x m^{(j)}_{x,i} \left(\hat n_x -  \langle \hat n_{x} \rangle^{(j)\prime}_i\right).
\end{align}
This expression is strictly valid for $\varphi \ll 1$, however, Ref.~\onlinecite{Altuntas2025} determined that in practice when $\varphi\lesssim 0.1$ statistical uncertainties will exceed systematic artifacts from higher order terms.
Thus, within this regime, the leading-order Kraus expansion accurately captures the backaction and measurement outcomes.

We now explicitly consider our experimental sequence starting with $\ket{\Psi_1^{(j)}}$ at $t_1=0$.
The measurement outcome and post-measurement state are
\begin{align}
\delta n^{(j)}_{x,1} &= \left(\langle \hat n_{x} \rangle^{(j)} - \langle \hat n_{x} \rangle\right) + \frac{m^{(j)}_{x,1}}{\varphi_1} & {\rm and} && \ket{\Psi_2^{(j)\prime}} &= \hat U\left[ 1 + \varphi_1 \sum_x m^{(j)}_{x,1} \left(\hat n_x -  \langle \hat n_{x} \rangle^{(j)} \right) \right]\ket{\Psi_1^{(j)}},
\end{align}
respectively.
The post-measurement state has already been advanced in time to $t_2=\delta t$ by the time evolution operator $\hat U$.
We adopt the short-hand $\langle\cdots\rangle = \langle\cdots\rangle_1$ since we will conclude with expectation values with respect to the initial state.

Likewise the noise in the second measurement is
\begin{align}
\delta n^{(j)}_{x,2} &= \left(\langle \hat n_{x} \rangle^{(j)\prime}_2 - \overline{\langle \hat n_{x} \rangle^{(j)\prime}}\right) + \frac{m^{(j)}_{x,2}}{\varphi_2}
\end{align}
The expectation value of the density  just prior to measurement $i=2$ is the most important quantity in all of these arguments:
\begin{align}
 \langle \hat n_{x} \rangle^{(j)\prime}_2 \approx \langle \hat n_{x,2} \rangle^{(j)} + \varphi_1\sum_{x^\prime} m^{(j)}_{x^\prime,1} \left[\langle \hat n_{x,2}\left(\hat n_{x^\prime,1} - \langle \hat n_{x^\prime,1} \rangle^{(j)}\right)\rangle^{(j)} +\ {\rm c.c}\right].
\end{align}
Notice that all expectation values on the right hand side are with respect to the initial state, and all time-evolution has been moved into the operators.
We now consider both terms of this expression.

First, averaging over trajectories quickly yields $\overline{\langle \hat n_{x} \rangle^{(j)\prime}_2} = \langle \hat n_{x,2} \rangle$, so the statistical contribution to the fluctuations will be
\begin{align}
\langle \hat n_{x} \rangle^{(j)\prime}_2 - \overline{\langle \hat n_{x} \rangle^{(j)\prime}} &= \langle \hat n_{x,2} \rangle^{(j)} - \langle \hat n_{x,2} \rangle \equiv \langle \delta \hat n_{x,2}\rangle^{(j)}.
\end{align}
Here we finally introduce the noise operator $\delta \hat n_{x,i} = \hat n_{x,i} - \langle\hat n_{x,i}\rangle=  \hat U_i^\dagger \hat n_{x} \hat U_i - \langle\hat U_i^\dagger \hat n_{x} \hat U_i\rangle$, which includes both statistical and quantum noise of the initial ensemble, but not projection noise from the measurement.

Next, the backaction contribution to the fluctuations requires more simplification
\begin{align}
\langle \hat n_{x,2}\left(\hat n_{x,1} - \langle \hat n_{x,1} \rangle^{(j)}\right)\rangle^{(j)} &= \langle \hat n_{x,2} \hat n_{x,1}\rangle^{(j)} - \langle \hat n_{x,2} \rangle^{(j)}\langle\hat n_{x,1}\rangle^{(j)} \\
&= \langle \delta\hat n_{x,2} \delta\hat n_{x,1}\rangle^{(j)} - \langle \delta\hat n_{x,2} \rangle^{(j)}\langle\delta\hat n_{x,1}\rangle^{(j)}
\end{align}
So assembling we find the relations for density and noise for both measurements
\begin{align}
n^{(j)}_{x,1} &= \langle \hat n_{x,1} \rangle^{(j)} + \frac{m^{(j)}_{x,1}}{\varphi_1} & {\rm and} && 
n^{(j)}_{x,2} &= \langle \hat n_{x,2}\rangle^{(j)} + \varphi_1 \sum_{x^\prime} m^{(j)}_{x^\prime,1} \left[ \langle \{  \hat n_{x,2}, \hat n_{x^\prime,1}\}\rangle^{(j)} - 2 \langle  \hat n_{x,2} \rangle^{(j)}\langle \hat n_{x^\prime,1}\rangle^{(j)}\right] + \frac{m^{(j)}_{x,2}}{\varphi_2}\\
\delta n^{(j)}_{x,1} &= \langle \delta \hat n_{x,1} \rangle^{(j)} + \frac{m^{(j)}_{x,1}}{\varphi_1} & {\rm and} &&
\delta n^{(j)}_{x,2} &= \langle \delta \hat n_{x,2}\rangle^{(j)} + \varphi_1 \sum_{x^\prime} m^{(j)}_{x^\prime,1} \left[ \langle \{ \delta\hat n_{x,2},\delta\hat n_{x^\prime,1}\}\rangle^{(j)} - 2 \langle \delta\hat n_{x,2} \rangle^{(j)}\langle\delta\hat n_{x^\prime,1}\rangle^{(j)}\right] + \frac{m^{(j)}_{x,2}}{\varphi_2}. \label{eqSM:measurements_x}
\end{align}
Since the density and noise versions look pretty much the same, we focus on the noise expressions in what follows.
In addition, notice that if every state in the ensemble is the same, then every $\langle \delta \hat n_{x,2}\rangle^{(j)}$ type term is zero, which greatly simplifies the noise analysis in that case.
Our analysis procedures identify various correlations---both as a function of $x$ and $\delta t$---present in back-to-back measurements.

For example, computing the spatially averaged CCF quickly leads to
\begin{align}
{\rm CCF}_{\delta x} = \frac{1}{N} \sum_x\overline{\delta n^{(j)}_{x,1}\delta n^{(j)}_{x + \delta x,2}} &= \frac{1}{2 N} \sum_x \langle \{ \delta\hat n_{x,1},\delta\hat n_{x + \delta x,2}\}\rangle,
\end{align}
having remembered that our random variable has variance of $1/2$.

%%%%%%%%%%%%%%%%% %%%%%%%%%%%%%%%%% %%%%%%%%%%%%%%%%% %%%%%%%%%%%%%%%%% 
\section{Weak values}\label{app:weak_values}
\ 
Here we describe an analysis methodology inspired by QWVs, a post-selection based measurement protocol introduced in 1988 by Aharonov and coauthors\cite{Aharonov1988}.
Aharonov's original protocol measures the same system twice: a weak measurement is followed immediately by a strong measurement of a different observable.
Then, each weak measurement outcomes is then accepted or rejected based on the corresponding strong measurement outcome.
The average over this post-selected ensemble is a QWV observable associated with the weakly measured quantity.
Here we develop an analogous strategy applied to our twice-weakly-measured system.

The essence of our strategy is most clearly understood for an ensemble with no statistical fluctuations, in which case Eqs.~\eqref{eqSM:measurements_x} reduce to
\begin{align}
\delta n^{(j)}_{x,1} &= \frac{m^{(j)}_{x,1}}{\varphi_1} & {\rm and} &&
\delta n^{(j)}_{x,2} &= \varphi_1 \sum_{x^\prime} m^{(j)}_{x^\prime,1} \langle \{ \delta\hat n_{x,2},\delta\hat n_{x^\prime,1}\}\rangle^{(j)} + \frac{m^{(j)}_{x,2}}{\varphi_2}.
\end{align}
We construct post-selected ensembles $\{ \delta n^{(j,\pm)}_{x,2} \}$ of second measurement outcomes for which $\pm \delta n^{(j)}_{x,1} > 0$.
Because $\varphi_1 >0$, this post selection process yields an ensemble of random variables $\{ \delta m^{(j,\pm)}_{x',1} \}$ with mean $\overline{m^{(j,\pm)}_{x',1}} = \pm\delta_{x,x^\prime} / \sqrt{\pi}$.

The average of the post-selected second measurement outcomes is therefore
\begin{align}
\overline{\delta n^{(j,\pm)}_{x,2}} &= \pm\frac{\varphi_{1}}{\sqrt{\pi}} \langle \{\delta\hat n_{x,1}^\dagger,\delta\hat n_{x,2}\}\rangle^{(j)},
\end{align}
clearly predicting that, unlike the cross-correlation approach, this weak-value signal is proportional to the measurement strength $\varphi_1$.

Notice that averages over the $\pm$ ensembles differ only by their overall sign.
In practice, we therefore use the signal $[\overline{\delta n^{(j,+)}_{x,2}} - \overline{\delta n^{(j,-)}_{x,2}}] / 2$; this includes the whole dataset in the analysis for reduced noise.

In the following section we consider the contribution of technical noise as well as a post-selection threshold other than zero.

\subsubsection*{Impact of technical noise}
Now we assume that each measurement of density has an additional noise source $q_{x,i}^{(j)}/\vartheta_i$, where $q_{x,i}^{(j)}$ is a random variable with variance $1/2$ and $\vartheta_i$ is an ``effective measurement strength''.
Since we considering post-selecting based on the sum of all noise sources in the first measurement, we require the mean of $m_{x,1}^{(j)}$ conditioned on $\delta n_{x,1}^{(j)} > 0$ (including the additive contribution of technical noise).

We therefore consider the joint probability distribution for the two contributions to the noise
\begin{align}
P(m,q) &= \frac{1}{\pi} \exp\left(-m^2 - q^2\right)
\end{align}
where for this part of the argument we omit the superscripts and subscripts.
The overall measurement noise $m / \varphi + q /\vartheta$ has zero mean, but now has variance $v = 1/(2 \varphi^2) + 1/(2 \vartheta^2)$: as usual, equal to the sum of the individual variances.
Now we compute the quantity of interest, the mean of $m$, conditioned on $m / \varphi + q /\vartheta > 0$.
The method of solution is to first create a new normalized conditional distribution
\begin{align}
P^\prime(m,q) &= 2 \begin{cases}
    P(m,q), & \text{if } m / \varphi + q /\vartheta > 0 \\
    0, & \text{if } m / \varphi + q /\vartheta \leq 0
\end{cases},
\end{align}
and from that evaluate the mean
\begin{align}
\overline m &= \frac{1}{\sqrt{\pi}} \frac{\vartheta}{\sqrt{\varphi^2 + \vartheta^2}} = \frac{1}{\sqrt{\pi}} \left(1 + \frac{\varphi^2}{\vartheta^2}\right)^{-1/2}
\end{align}
The important aspect of this result is that for the beam splitter part of the model, the added noise is a fixed fraction of the measurement noise, so this attenuation factor is a constant.  Also in the limit of small $\vartheta$ (large added noise) we arrive at $\overline m \approx \vartheta / (\sqrt{\pi} \varphi)$.

Next we consider the case of a variable cutoff $T$ (i.e. threshold) measured as a fraction of the width of the overall distribution, i.e., we now keep data where
\begin{align}
\frac{m}{\varphi} + \frac{q}{\vartheta} > T \sqrt{v}.
\end{align}
In looking at the resulting integrals, it becomes clear that it is better to work in a rotated coordinate system with
\begin{align}
m &= m' \cos \theta - q' \sin \theta & {\rm and} && q &= m' \sin \theta + q' \cos \theta
\end{align}
giving the observed degree of freedom 
\begin{align}
\frac{m}{\varphi} + \frac{q}{\vartheta} &= \frac{m' \cos \theta - q' \sin \theta}{\varphi} + \frac{m' \sin \theta + q' \cos \theta}{\vartheta}\\ 
&= m' \left(\frac{\cos \theta}{\varphi} + \frac{\sin \theta}{\vartheta} \right) + q' \left(\frac{-\sin \theta}{\varphi} + \frac{\cos \theta}{\vartheta} \right) \\
 &= m' \sqrt{2v}.
\end{align}
In the last line we identified the condition $\tan\theta = \varphi/\vartheta$ where $q^\prime=0$.

This implies that the cutoff takes the simple form
\begin{align}
m' &> \frac{T}{\sqrt{2}} & \text{giving retained fraction}  && r &= \frac{1}{2}{\rm erfc}\left(\frac{T}{\sqrt{2}}\right),
\end{align}
where ${\rm erfc}(\cdots)$ is the complementary error function.
Computing leads to the final result
\begin{align}
\overline m(T) &= \frac{1}{\sqrt{\pi}} \cos\theta \left[\frac{e^{-T^2 / 2}}{{\rm erfc}(T/\sqrt{2})} \right] = \overline m \left[\frac{e^{-T^2 / 2}}{{\rm erfc}(T/\sqrt{2})} \right]\\
& = \overline m \left[\frac{e^{-[{\rm erfc}^{-1}\left(2 r\right)]^2}}{2 r}\right],
\end{align} 
keeping in mind that here $r$ is the positive only retained fraction, so $f = 1-2r$ is what we consider as the ``fraction discarded'' in the main body.

%%%%%%%%%%%%%%%%% %%%%%%%%%%%%%%%%% %%%%%%%%%%%%%%%%% %%%%%%%%%%%%%%%%% 
\section{Resolution limited measurement strength}\label{app:strength}
\ 
In the main body of the text we introduced two metrics of measurement strength $\varphi$, and $g$.
The first relates to photon shot noise on an imaging sensor (quantum projection noise associated with measuring the photon number of a coherent state of light), and the second derives from the atomic scattering rate $N_{\rm scat}^{1/2}$.
Here we establish the relationship between these quantities.

The PCI signal $g_{\rm PCI}$ in Eq.~\eqref{eq:PCI} can be reexpressed as $g_{\rm PCI} = 1 - N / N_0$, in terms of the per-pixel number of photons with no atoms present $N_0$, and the number with atoms (and PCI phase dot) present $N$.
In the small signal limit (with $N \lesssim N_0$), the PCI uncertainty is $\Delta g_{\rm PCI} \approx N_0^{-1/2}$.
For a sensor with pixel of area $A$, the initial photon number is $N_0 = I A \tm / (\hbar \omega_0)$ and the deduced atom number and uncertainty are
\begin{align}
n &= 2\frac{\delta}{\Gamma} \left(\frac{A}{\sigma_0}\right)  g_{\rm PCI}  & {\rm and} && \Delta n &= \frac{2}{N_0^{1/2}}\frac{\delta}{\Gamma} \left(\frac{A}{\sigma_0}\right),
\end{align}
where  $\sigma_0 = 6 \pi / k_0^2$ is the resonant scattering cross-section.
The uncertainty can be related to $N_{\rm scat}$ via
\begin{align}
\Delta n &= 2(\dbar)\left(\frac{\hbar \omega_0}{I \tm}\right)^{1/2} \left(\frac{A^{1/2}}{\sigma_0}\right)\\
&= 2 \left(\frac{(\dbar)}{[(\Ibar) (\tbar)]^{1/2}}\right) \left(\frac{\hbar \omega_0 \Gamma}{\Isat}\right)^{1/2} \left(\frac{A^{1/2}}{\sigma_0}\right) \\
\Delta n &= 2\sqrt{2} \left(\frac{(\dbar)^2}{(\Ibar)(\tbar)}\right)^{1/2} \left(\frac{A}{\sigma_0}\right)^{1/2} = \left(\frac{A}{N_{\rm scat} \sigma_0}\right)^{1/2},
\end{align}
in terms of the mean number of scattered photons per atom $N_{\rm scat} = (\tbar)\,(\Ibar) / [8 (\dbar)^2]$.
In our past experimental work\cite{Altuntas2023a,Altuntas2023,Altuntas2023b}, we used the simple dimensionless ratio $g = \sqrt{\Ibar \tbar} / \dbar$ to quantify measurement strength, giving $N_{\rm scat} = g^2/8$.

On the other hand in our formal description, the pixel-by-pixel measurement outcomes are described by the relations
\begin{align}
n_{\bf r} & = \langle\hat n_{\bf r}(t)\rangle + \delta n_{\bf r}, & \delta n_{\bf r} &= \frac{m_{\bf r}}{\varphi}, & {\rm and} && \overline{m_{\bf r} m_{{\bf r}^\prime}} &= \frac{\delta_{{\bf r},{\bf r}^\prime}}{2},
\end{align}
giving a per-pixel number uncertainty of $1/(\sqrt{2} \varphi)$.
Equating these two expressions for number uncertainty yields
\begin{align}
\varphi &= \left(\frac{N_{\rm scat} \sigma_0}{2 A}\right)^{1/2} = \frac{g}{4} \left(\frac{\sigma_0}{A}\right)^{1/2},
\end{align}
and as a result linking the strength $N_{\rm scat}^{1/2}$ suitable for scattering measurements to $\varphi$ for imaging experiments.

These relations can also be expressed using momentum-space variables.
First, observe that the $k$-space window derived from a 2D image with pixel size $A$ has area $(2 \pi)^2 / A$ in which detection noise is uniformly distributed.
By contrast, the resolution-limit might alternately be limited by $k_{\rm NA}$, the maximum transverse wavevector accepted by the imaging system (or the numerical aperture ${\rm NA} = k_{\rm NA} / k_0$).
In this case, detection noise is uniformly distributed in $k$-space disk with area $\pi k_{\rm NA}^2$.
Equating these two different momentum-space areas yields
\begin{align}
\varphi &= \frac{3^{1/2}}{2}  N_{\rm scat}^{1/2}\times {\rm NA} = \frac{1}{4} \left(\frac{3}{2}\right)^{1/2} g \times {\rm NA}.
\end{align}
As a sanity check, we can take ``nature's resolution'' with $ k_{\rm NA} = k_0$ to find $\varphi \simeq \sqrt{3} N_{\rm scat}^{1/2}/2 = 0.87 N_{\rm scat}^{1/2}$.  
While specifics are hand-wavy, the basic point is that these two metrics of measurement strength are essentially the same at their fundamental limit.
For our strongest measurements, with $N_{\rm scat}^{1/2} \approx 0.125$ (i.e., $g \approx 1.0$), we estimate $\varphi \approx 0.1$: well below unity, ensuring that the measurement remains in the weak limit where higher-order terms are negligible.

%%%%%%%%%%%%%%%%% %%%%%%%%%%%%%%%%% %%%%%%%%%%%%%%%%% %%%%%%%%%%%%%%%%% 
\section{Finite temperature speed of sound}\label{app:speed_of_sound}
\ 
Here we develop a simple model describing the speed of sound of a finite-temperature harmonically trapped BEC.
Unlike for uniform confinement, i.e. a box trap, when the condensate fraction falls, the majority of thermal excitations reside spatially outside of the condensate mode.
Hence, our model assumes leading order effect of $R_c < 1$ is to reduce the number of atoms in the BEC, thereby reducing the chemical potential and sound speed.
In our laboratory implementation, we increased the temperature at fixed atom number by simultaneously reducing the initial atom number and increasing depth of the dipole trap.
This lead to an increase in trap frequency that is included in our modeling.

We begin with the total atom number of a BEC in the Thomas-Fermi (TF) approximation
\begin{align}
N &= \frac{1}{15} \left(\frac{\ell_{\rm ho}}{a}\right)\left(\frac{2\mu}{\hbar\bar \omega}\right)^{5/2},
\end{align}
where $\bar\omega^3 \equiv \omega_x \omega_y \omega_z$ is the geometric average of the trap frequencies; $a$ is the scattering length; $\ell_{\rm ho} = ({\hbar} / {m \bar\omega})^{1/2}$ is the harmonic oscillator length; and $\mu$ is the chemical potential.
This relationship asserts the scaling between speed of sound and the total atom number is $c \propto \mu^{1/2} \propto N^{1/5}$; accounting for the trap frequency correction this changes to $c \propto N^{1/5} \bar\omega^{3/5}$.
Next, normalizing by the low-temperature speed of sound $c_0$ we obtain 
\begin{align}
\frac{c}{c_0} &= R_c^{1/5} \left(\frac{\omega^2}{\omega_0^2}\right)^{3/10}\label{eq:app:c_over_c0},
\end{align}
where $\omega_0$ is the trap frequency as $R_c \rightarrow 1$. 
In anticipation of the $\omega \propto P^{1/2}$ scaling with laser power $P$, we expressed the ratio of the trap frequencies in terms of $\omega^2$. 

Utilizing time-of-flight absorption imaging measurements, we empirically found the condensate fraction is a linear function $R_c = (P_c - P) / (P_c - P_0)$ of laser power, where $P_0$ and $P_c$ denote the laser powers for $R_c\rightarrow 1$ and $R_c\rightarrow 0$, respectively.
Rearranging to parallel Eq.~\eqref{eq:app:c_over_c0} yields
\begin{align}
\frac{P}{P_0} &= \frac{P_c}{P_0} + R_c \left(1 - \frac{P_c} {P_0}\right)  &\rightarrow 
&& \frac{\omega^2}{\omega_0^2} &= \frac{\omega_c^2}{\omega_0^2} + R_c \left(1 - \frac{\omega_c^2}{\omega_0^2}\right).
\end{align}
Finally, incorporating into Eq.~\eqref{eq:app:c_over_c0} yields the speed of sound of a finite-temperature harmonically trapped BEC
\begin{align}
\frac{c}{c_0} &= R_c^{1/5} \left[\frac{\omega_c^2}{\omega_0^2} + R_c \left(1 - \frac{\omega_c^2}{\omega_0^2}\right) \right]^{3/10}.
\end{align}
This expression breaks down near $T_c$, when a significant thermal fraction spatially overlaps with the condensate mode.
In Fig.~4 and surrounding text of the main manuscript, we use this model to benchmark the dependence of the Van Hove function on the condensate fraction.

%TC:endignore

% \appendix

\end{document}